\documentclass[aps,prb,longbibliography,twocolumn,showpacs,superscriptaddress]{revtex4-1}
\usepackage{amssymb}
\usepackage{amsbsy}
\usepackage{amsmath}
\usepackage{enumitem}
\usepackage[pdftex]{graphics}
\usepackage{color}
\usepackage{float}
\usepackage{comment}
\usepackage{graphicx}
\usepackage{dcolumn}
\usepackage{subfigure}
\usepackage{bm}
\usepackage{natbib}
\def\be{\begin{equation}}
\def\ee{\end{equation}}
\def\beq{\begin{equation}}
\def\eeq{\end{equation}}

\usepackage{bm}        
\usepackage{amssymb}   
\usepackage{exscale}
\usepackage{xcolor}
\newcommand{\ket}[1]{|#1\rangle}

\newcommand{\marcos}[1]{\textcolor{black}{ #1}}

\newcommand{\blue}[1]{\textcolor{black}{ #1}}

\definecolor{darkred}{rgb}{0.90,0.2,0.2}

\usepackage[plainpages=false,pdfpagelabels,colorlinks=true,linkcolor=red,urlcolor=blue,citecolor=blue,pdftitle={},pdfauthor={},pdfdisplaydoctitle={},pdfduplex=DuplexFlipLongEdge]{hyperref}

\begin{document}

\title{Signatures of quantum phase transitions after quenches in\\ quantum-chaotic one-dimensional systems}

\author{Asmi Haldar}
\affiliation{Indian Association for the Cultivation of Science (School of Physical Sciences), 2A \& 2B Raja S.~C.~Mullick Road, Kolkata 700032, India}
\affiliation{Max Planck Institute for the Physics of Complex Systems, N{\"o}thnitzer Stra{\ss}e 38, D-01187, Dresden, Germany}

\author{Krishnanand Mallayya}
\affiliation{Department of Physics, The Pennsylvania State University, University Park, Pennsylvania 16802, USA}
\affiliation{Department of Physics, Cornell University, Ithaca, New York 14850, USA}

\author{Markus Heyl}
\affiliation{Max Planck Institute for the Physics of Complex Systems, N{\"o}thnitzer Stra{\ss}e 38, D-01187, Dresden, Germany}

\author{Frank Pollmann}
\affiliation{Department of Physics, Technische Universit\"{a}t M\"{u}nchen, James-Franck-Str.~1/I 85748 Garching b.~M\"{u}nchen, Germany}

\author{Marcos Rigol}
\affiliation{Department of Physics, The Pennsylvania State University, University Park, Pennsylvania 16802, USA}

\author{Arnab Das}
\affiliation{Indian Association for the Cultivation of Science (School of Physical Sciences), 2A \& 2B Raja S.~C.~Mullick Road, Kolkata 700032, India}

\begin{abstract}
Quantum phase transitions are central to our understanding of why matter at very low temperatures can exhibit starkly different properties upon small changes of microscopic parameters. Accurately locating those transitions is challenging experimentally and theoretically. Here we show that the antithetic strategy of forcing systems out of equilibrium via sudden quenches provides a route to locate quantum phase transitions. Specifically, we show that such transitions imprint distinctive features in the intermediate-time dynamics, and results after equilibration, of local observables in quantum-chaotic spin chains. Furthermore, we show that the effective temperature in the expected thermal-like states after equilibration can exhibit minima in the vicinity of the quantum critical points. We discuss how to test our results in experiments with Rydberg atoms, and explore nonequilibrium signatures of quantum critical points in models with topological transitions.
\end{abstract}
    
\maketitle

\section{Introduction}\label{Intro}

Quantum phase transitions are key to our perception of quantum matter across fields in physics, from quark-gluon plasma and neutron stars to quantum magnets and high-temperature superconductors~\cite{Subir-Book, Carr_QPT_Book}. At those transitions different quantities in completely different systems can exhibit universal behavior. This is something that we understand thanks to the development of the renormalization group theory. Among the challenges that remain for each specific system is to (if possible) find experimentally where quantum phase transitions occur, as well as theoretically predict their locations using simplified models. Quantum simulators promise to overcome the latter challenge by experimental means, as they provide pristine and controllable realizations of theoretical models~\cite{bloch_dalibard_review_08, esslinger_review_10}. 

Quantum simulators also provide access to real-time dynamics. This is something that can be used to explore unique aspects of crossing a quantum phase transition in real time. For example, recently a Rydberg-atom quantum simulator was used to probe the Kibble-Zurek mechanism of universal defect production for slow parameter sweeps~\cite{2019Keesling}. On the theoretical side, recent works have provided evidence that nonequilibrium quantum evolution can be used to probe quantum phase transitions in integrable systems~\cite{SciRep_BDD, Topological_QPTSgn_RMD}, in prethermal states for models close to integrability~\cite{Gong_Titum}, or through out-of-time-order correlators~\cite{2018Heyl}. However, identifying real-time signatures of quantum phase transitions in generic (quantum-chaotic) many-body systems has remained a challenge.

In this work we show that generic quantum matter can exhibit dynamical signatures of quantum phase transitions by the antithetic strategy of forcing these systems out of equilibrium and therefore beyond the ground-state manifold. We find that the intermediate-time dynamics of local observables and of the entanglement entropy exhibit distinct features after quantum quenches in the anisotropic next-nearest neighbor Ising (ANNNI) chain upon tuning the quench parameter across an underlying quantum phase transition. Specifically, we find that the derivatives of local observables with respect to the quench parameter develop prominent dips/peaks in the vicinity of the quantum phase transition. We determine the quantum real-time evolution by means of the infinite-Time Evolved Block Decimation (iTEBD), which provides numerically exact results for the transient to intermediate-time dynamics in the thermodynamic limit~\cite{Vidal_1, *Vidal_2, *Vidal_3, Schlwck_dmrg_mps, Hauschild_Frank_iTEBD}. 

In order to access the long-time (asymptotic) properties of the considered quantum-chaotic system after the expected thermalization, we employ a numerical linked cluster expansion (NLCE) for thermal equilibrium states~\cite{rigol2006numerical, *rigol2007numerical}. We again find distinct signatures of the quantum phase transition in derivatives of the correlation functions. Also, the effective temperature exhibits a marked minimum as function of the quench parameter in the close vicinity of the quantum phase transition. Since the considered one-dimensional system does not support singular behavior after equilibration, upon assuming that eigenstate thermalization occurs~\cite{deutsch_91, srednicki_94, rigol2008thermalization, dalessio_kafri_16}, these prominent features are not associated with nonanalytic properties (in contrast to the integrable systems studied in Refs.~\cite{SciRep_BDD, Topological_QPTSgn_RMD}), but nevertheless represent distinct signatures of quantum phase transitions. Finally, we discuss similar phenomena for quantum phase transitions involving topologically different quantum states. We also discuss how our findings can be tested in current experiments with Rydberg atoms.

The presentation is organized as follows. In Sec.~\ref{sec:Protocol}, we introduce the Hamiltonian of the ANNNI chain, and introduce the protocol used to probe the ferromagnetic to paramagnetic quantum phase transition via dynamics following quantum quenches. The results obtained for dynamics after the quenches are presented in Sec.~\ref{sec:dynamics}, while the results after thermalization are presented in Sec.~\ref{sec:thermalization}. Combining results from the dynamics and thermalization, in Sec.~\ref{sec:phasediagram} we report the estimated phase diagram for the ferromagnetic to paramagnetic quantum phase transition for a wide range of parameters of the ANNNI chain. In Sec.~\ref{sec:experiments}, we discuss the feasibility of testing our results experimentally, while in Sec.~\ref{sec:topological} we discuss the applicability of our protocol to detect topological quantum phase transitions. In Sec.~\ref{sec:discussion}, we summarize our results and discuss their implications.
       
\section{ANNNI Hamiltonian and\\ Quench Protocol}
\label{sec:Protocol}

The ANNNI chain is a very well studied spin model (see, e.g., Ref.~\cite{BKC-Book}). Its Hamiltonian in a chain with $L$ sites can be written as 
\beq\label{H_ANNNI}
\hat H \doteq -\sum_{i}^{L}\sigma_{i}^{x}\sigma_{i+1}^{x} +\kappa\sum_{i}^{L}\sigma_{i}^{x}\sigma_{i+2}^{x} -\Gamma\sum_{i}^{L}\sigma_{i}^{z}.
\eeq
When mapped onto a fermionic Hamiltonian using Jordan-Wigner transformation~\cite{cazalilla_citro_review_11}, the next-nearest-neighbor term (with strength $\kappa$) maps onto a four-fermion interaction. At $T=0$, this model has a rich (and still partly controversial) phase diagram in the $\kappa$-$\Gamma$ plane. The quantum phase transition line from the ferromagnetic to the paramagnetic phase, which occurs as the antiferromagnetic next-nearest neighbor coupling $\kappa > 0$ crosses a critical value for a fixed $|\Gamma| < 1$, is a second order phase transition (see, e.g., Ref.~\cite{ANNNI_Line}). In the quadrant ($\kappa>0$, $\Gamma>0$), this line is well described using second order perturbation theory, with the critical parameters satisfying (see Fig.~\ref{fig:phase_obs})~\cite{ANNNI_Line}:
\beq\label{eq:ANNNI_Line}
1 - 2\kappa_c = \Gamma_c - \Gamma_c^{2}\frac{\kappa_c}{2(1-\kappa_c)}.
\eeq

\begin{figure}[!t]
\includegraphics[width=0.85\linewidth]{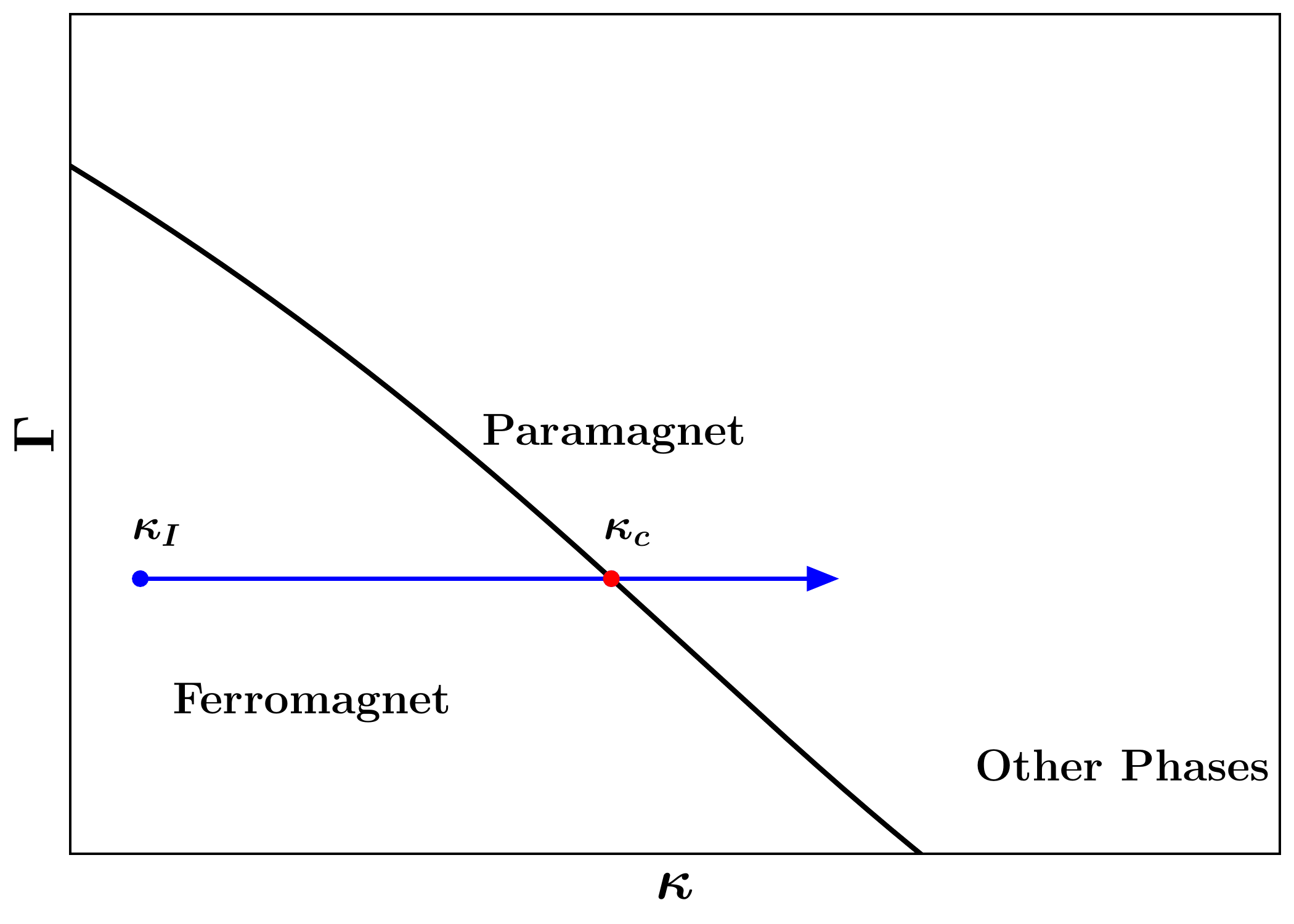}
  \caption{Schematic representation of our quench protocol, superimposed on a schematic ground-state phase diagram of the ANNNI chain.}
\label{Schematic}
\end{figure}

To probe this ferromagnetic to paramagnetic quantum phase transition at a fixed value of $\Gamma$, we generate a family of Hamiltonians $\hat H(\kappa)$. We then generate a family of nonequilibrium states via quenches with $\hat H(\kappa)$. The protocol (straightforward to generalize to other models) consists of following steps (see Fig.~\ref{Schematic}):

\begin{enumerate}[label=(\roman*), itemsep=1pt, topsep=1pt]
  \item The initial state is fixed to be the ground state $|\psi(\kappa_{I})\rangle$ of $\hat H(\kappa_{I})$, where $\kappa_{I}$ is deep in the ferromagnetic phase. 

  \item We suddenly change (quench) $\kappa_{I}\rightarrow\kappa$ at $t=0$, and study the unitary time evolution of the system under the time-independent Hamiltonian $\hat H(\kappa)$, i.e., $|\psi(t,\kappa)\rangle=\exp[-i\hat H(\kappa)t]|\psi(\kappa_{I})\rangle$ (we set $\hbar=1$).

  \item We compute expectation values of observables ${\cal O}(t,\kappa)=\langle\psi(t,\kappa)|\hat{\cal O}|\psi(t,\kappa)\rangle$.

  \item For a fixed value of $t$, we study how ${\cal O}(t,\kappa)$ changes with $\kappa$, focusing on the behavior in the vicinity of $\kappa_c$, where $\kappa_c$ is the critical value of $\kappa$ for the transition given the selected value of $\Gamma$.
\end{enumerate}

\section{Quantum Dynamics}\label{sec:dynamics}

We study the time evolution of observables after quantum quenches in an infinite ANNNI chain using iTEBD (see Appendix~\ref{Apndx:comp})~\cite{Vidal_1, *Vidal_2, *Vidal_3, Schlwck_dmrg_mps, Hauschild_Frank_iTEBD}. Following the protocol introduced in Sec.~\ref{sec:Protocol} (see Fig.~\ref{Schematic}), we fix $\Gamma$ (we take $\Gamma=0.2$) and then fix $\kappa_{I}$ so that the initial state is a ground state of the ANNNI chain deep in the ferromagnetic phase (we take $\kappa_{I}=0$). In our iTEBD calculations, we introduce a very small ($\sim 10^{-6}$) longitudinal field to pin one of the two degenerate maximally polarized ground states. The critical value of $\kappa$ for the ferromagnetic to paramagnetic quantum phase transition for $\Gamma=0.2$ is $\kappa_c\approx0.41$.

We first focus on the dynamics of two local observables, the nearest and next nearest neighbor longitudinal correlators
\beq\label{Cx}
C^{x}_{1(2)} = \frac{1}{L}\sum_{i=1}^{L}\langle\sigma^{x}_{i}\sigma_{i+1(2)}^{x}\rangle. 
\eeq

\begin{figure}[!t]
\includegraphics[width=0.99\linewidth]{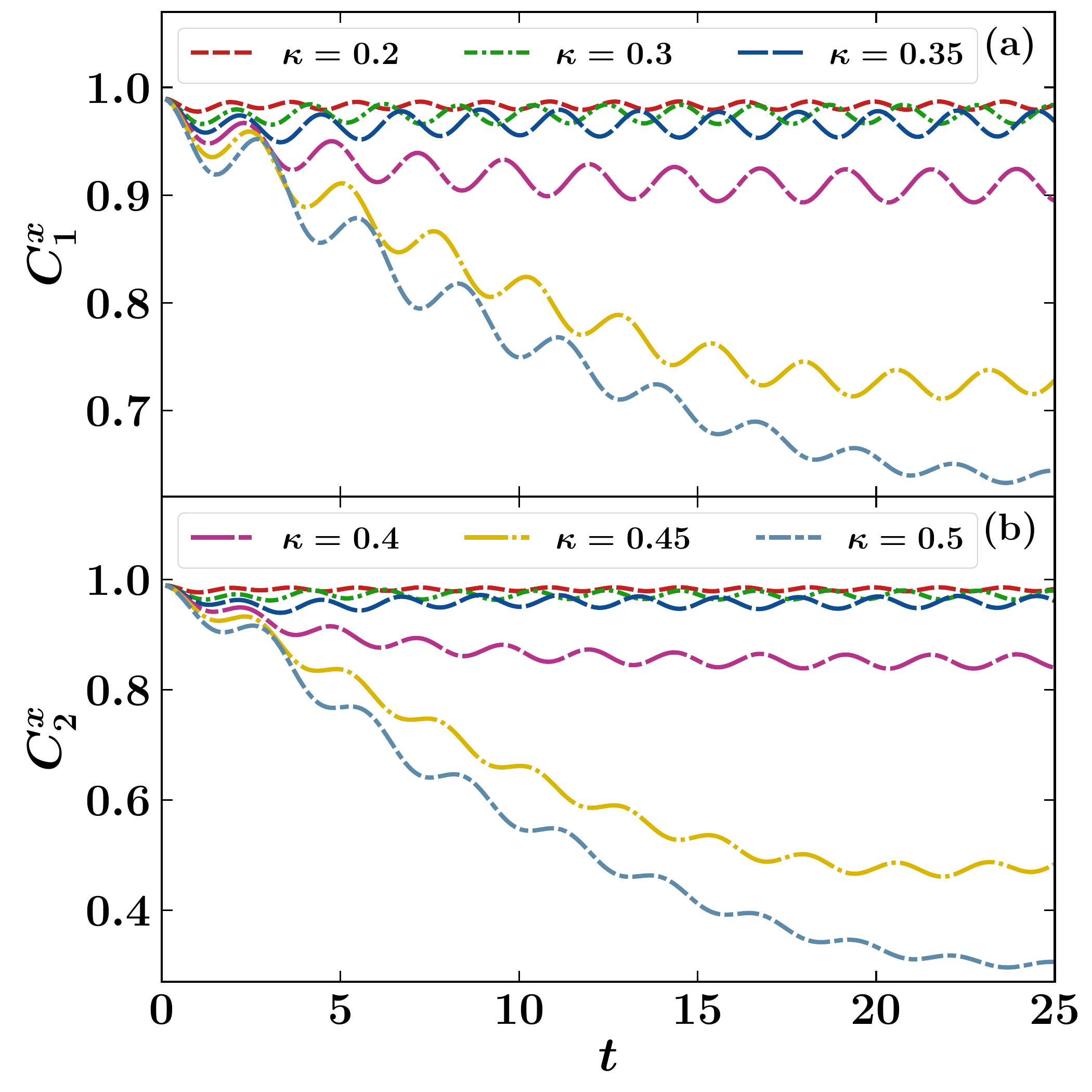}
  \caption{Time evolution of (a) $C^{x}_{1}$ and (b) $C^{x}_{2}$ for six values of $\kappa$ after quenches starting from the ground state of the ANNNI Hamiltonian with $\Gamma=0.2$ and $\kappa_I=0$.}
\label{Cx_time}
\end{figure}
\begin{figure}[!b]
\includegraphics[width=0.99\linewidth]{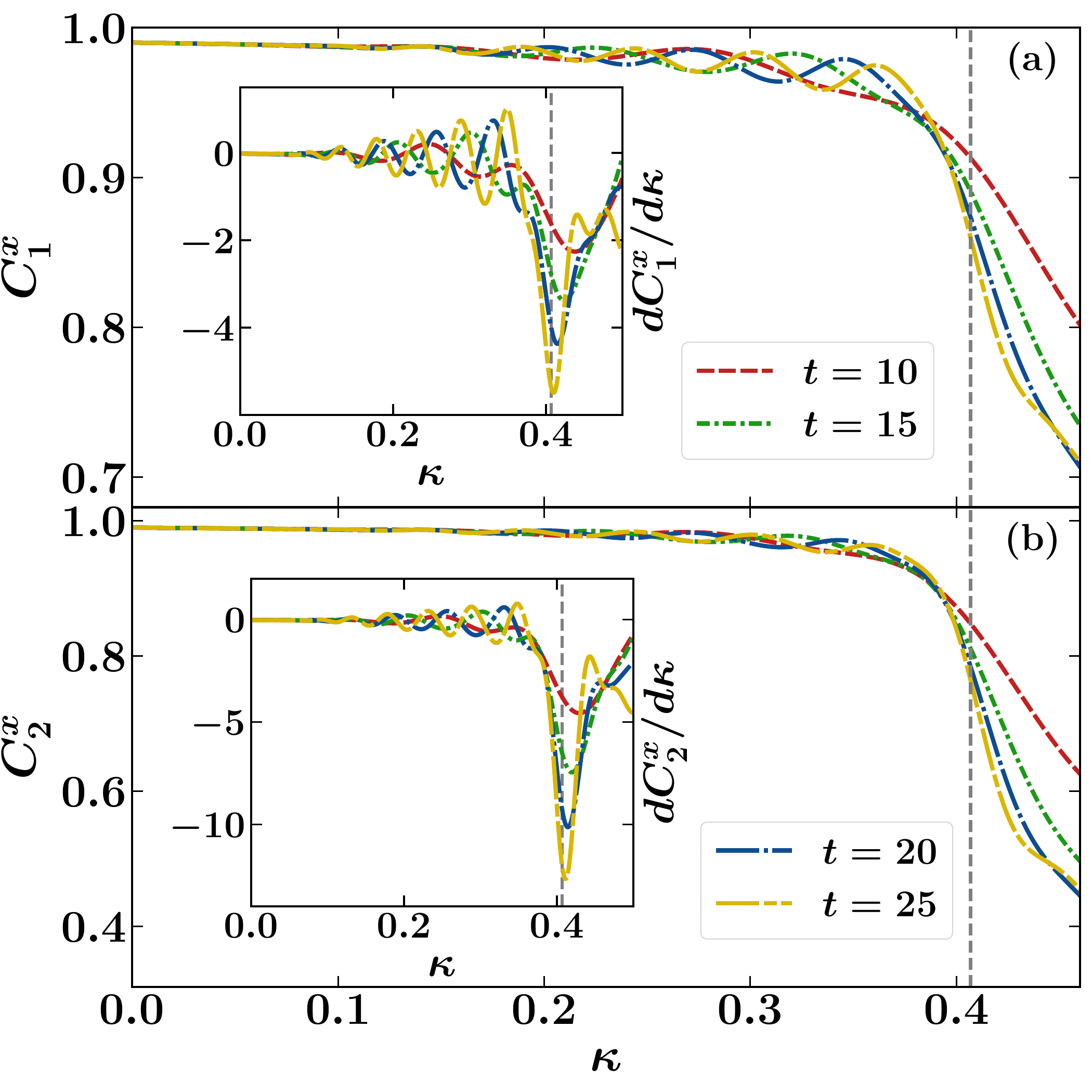}
  \caption{Ferromagnetic to paramagnetic quantum phase transition in the ANNNI chain as revealed via real time dynamics of local observables. (a) $C^{x}_{1}$ and (b) $C^{x}_{2}$ at different times plotted as functions of $\kappa$ after the quench (the legend is the same for both observables). (Insets) Derivative with respect to $\kappa$ of the results shown in the main panels. The initial state is the ground state of the ANNNI Hamiltonian with $\Gamma=0.2$ and $\kappa_I=0$. The vertical dashed lines mark the critical $\kappa_c\approx0.41$.}
\label{Cx_PB}
\end{figure}

In Fig.~\ref{Cx_time}, we show results for the time evolution of $C^{x}_{1}$ [Fig.~\ref{Cx_time}(a)] and $C^{x}_{2}$ [Fig.~\ref{Cx_time}(b)] for six values of $\kappa$ after the quench. The dynamics of both longitudinal correlators is qualitatively similar for the values of $\kappa$ shown. Their decrease with time speeds up as $\kappa$ increases about $\kappa_c$. How the closeness to $\kappa_c$ affects the dynamics is better seen by plotting the correlations for fixed times $t$ after the quench as functions of $\kappa$ [step (iv) in the protocol introduced in Sec.~\ref{sec:Protocol}]. This is done in Fig.~\ref{Cx_PB}, where we show results for $C^{x}_{1}$ [Fig.~\ref{Cx_PB}(a)] and $C^{x}_{2}$ [Fig.~\ref{Cx_PB}(b)]. At all times reported, $C^{x}_{1}$ and $C^{x}_{2}$ decrease rapidly with increasing the value of $\kappa$ for $\kappa\gtrsim\kappa_c$. In addition, with increasing time, the decrease in the correlators becomes more prominent when $\kappa\approx\kappa_c$. This is apparent in the insets, where we show the derivative of the correlators. They develop sharper dips close to $\kappa_c$ as the evolution time increases.

\begin{figure}[!b]
\includegraphics[width=0.99\linewidth]{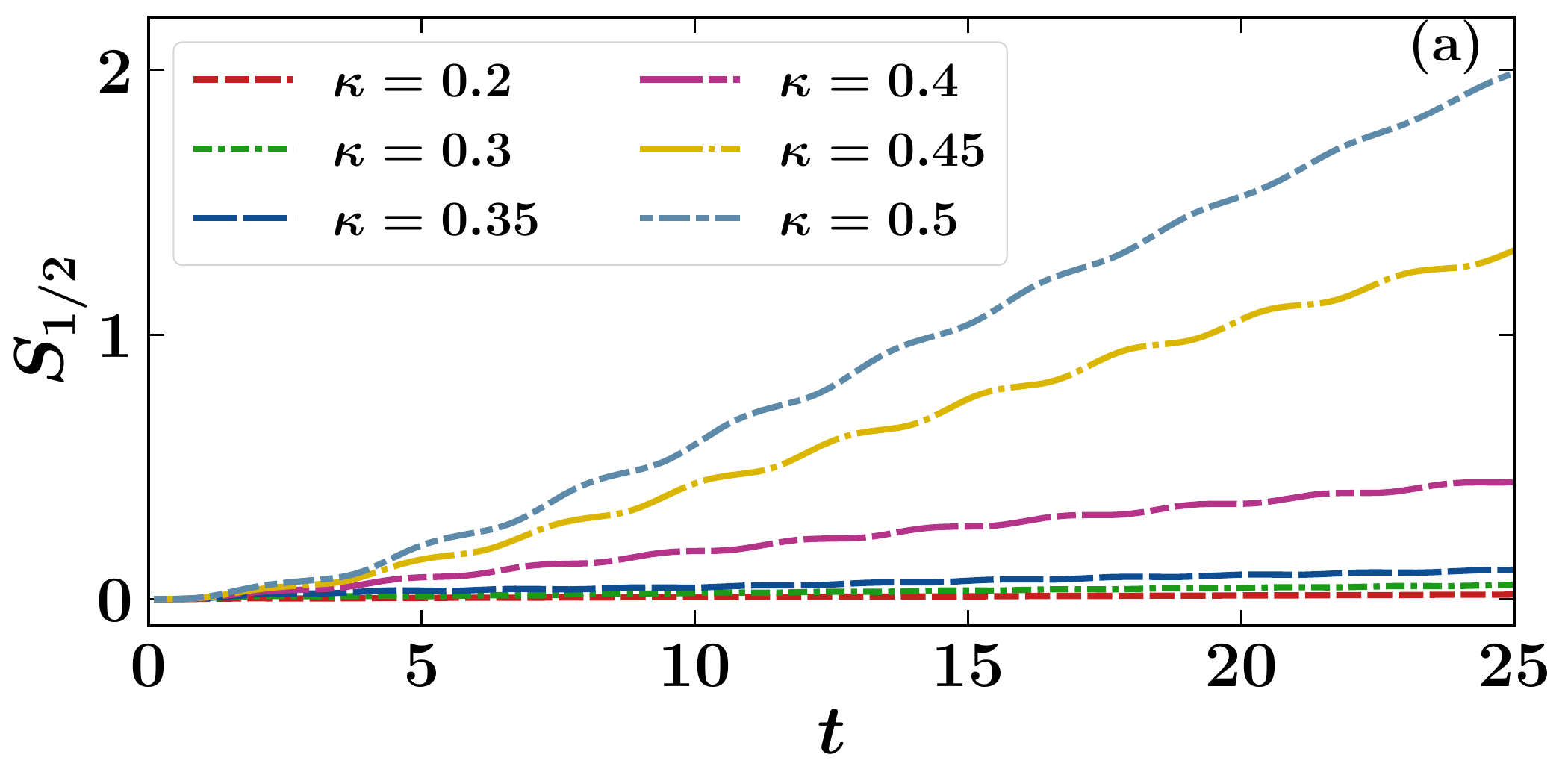}
\includegraphics[width=0.99\linewidth]{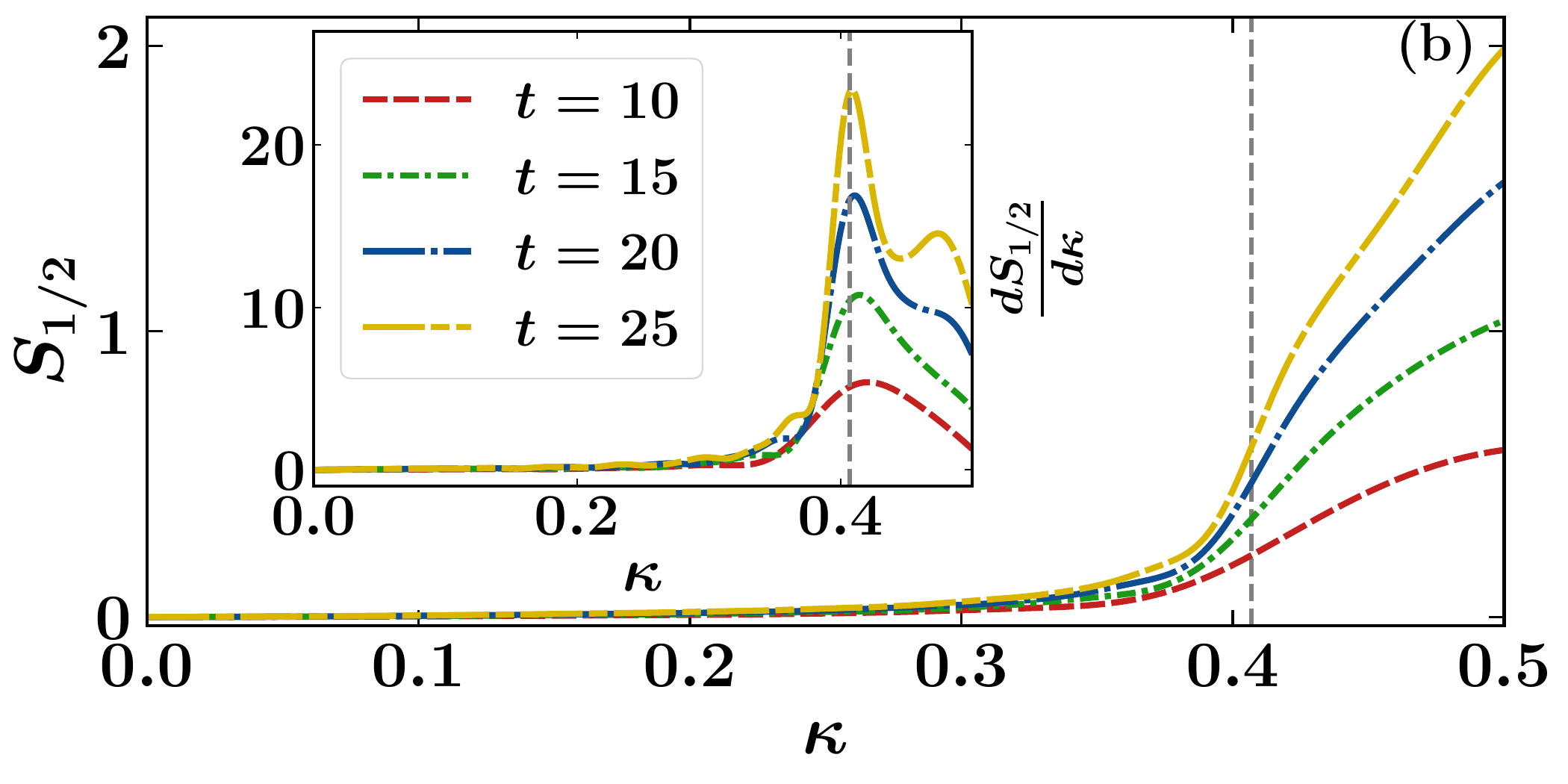}
  \caption{Ferromagnetic to paramagnetic quantum phase transition in the ANNNI chain as revealed via real time dynamics of the half-chain entanglement entropy $S_{1/2}$. (a) Time evolution of $S_{1/2}$ for six values of $\kappa$. (b) $S_{1/2}$ at different times plotted as a function of $\kappa$. (Inset) Derivative with respect to $\kappa$ of the results shown in the main panel. The initial state for the dynamics is the ground state of the ANNNI Hamiltonian with $\Gamma=0.2$ and $\kappa_I=0$.  The vertical dashed lines mark the critical $\kappa_c\approx0.41$.}
\label{Others}
\end{figure}

In Refs.~\cite{SciRep_BDD, Topological_QPTSgn_RMD} it was proved that following the same protocol discussed here but for noninteracting models (or models mappable to them) results in nonanalytic behavior of local observables at the quantum phase transition in the limit $t\rightarrow\infty$ (after having taken the thermodynamic limit first). While this is not the case in the quenches in generic models studied here (see Sec.~\ref{sec:thermalization}), the prominent features seen in Fig.~\ref{Cx_PB} at finite times are promising for an experimental determination of $\kappa_c$.

We also studied the dynamics of the half chain entanglement entropy $S_{1/2} = -\text{Tr}{[\rho_{1/2} \ln{\rho_{1/2}}]}$, where $\rho_{1/2}$ is the density matrix of the half chain (obtained by tracing out the other half). This is a nonlocal observable that is expected to increase linearly with time in quantum-chaotic systems~\cite{kim_huse_13}. In Fig.~\ref{Others}(a), we plot the time evolution of $S_{1/2}$ for six values of $\kappa$. As for the local operators in Fig.~\ref{Cx_time}(a), the change of $S_{1/2}$ with time speeds up as $\kappa$ increases about $\kappa_c$. Figure~\ref{Others}(b) shows $S_{1/2}$ at fixed times $t$ after the quench plotted vs $\kappa$, and the inset in Fig.~\ref{Others}(b) shows the derivative with respect to $\kappa$ of the results in the main panel. Like the local operators in Fig.~\ref{Cx_PB}, the behavior of the half chain entanglement entropy carries a marker of the quantum phase transition.

\subsection{Changing $\kappa_I$}\label{sec:timekappa_I}

\begin{figure}[!h]
\includegraphics[width=0.99\linewidth]{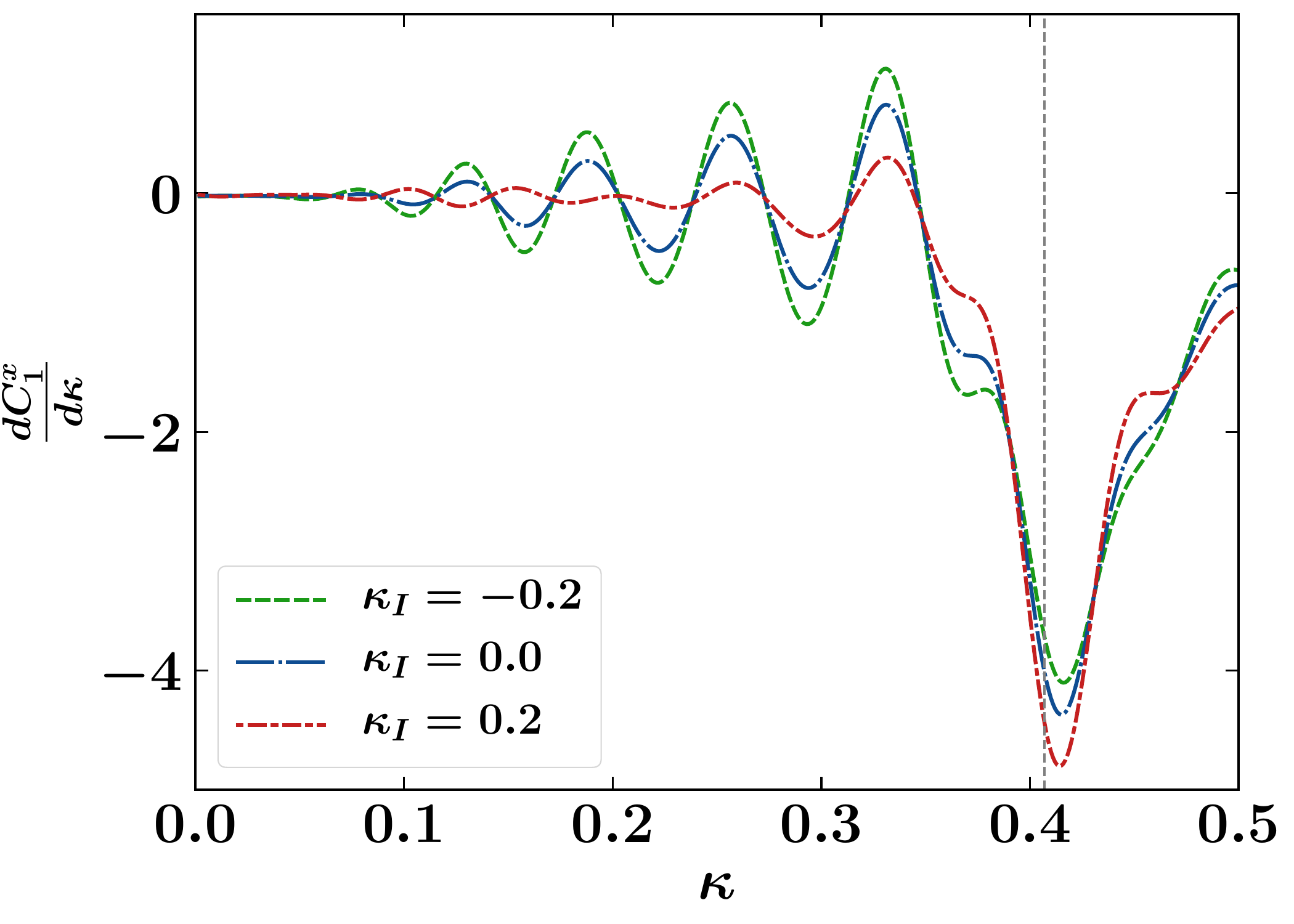}
  \caption{Results for $dC^{x}_{1}/d\kappa$, as those in the inset in Fig.~\ref{Cx_PB}(a), obtained at a fixed time $t=15$ after the quench for different values of $\kappa_{I}$ (=-0.2, 0, 0.2) in the initial ground state. The vertical dashed line marks the critical $\kappa_c\approx0.41$.}
\label{Pulse_Height_2}
\end{figure}

In Fig.~\ref{Pulse_Height_2}, we show the derivative of $C^{x}_{1}$ with respect to $\kappa$ at a fixed time after the quench, plotted as a function of $\kappa$, for different values of $\kappa_{I}$ in the initial ground state. We recall that as $\kappa_{I}$ departs from $\kappa_c$, for $\kappa_{I}<\kappa_c$, the ground state of the system is deeper in the ferromagnetic phase. The results in Fig.~\ref{Pulse_Height_2} show that starting deeper in the ferromagnetic phase results in a slightly shallower dip in $dC^{x}_{1}/d\kappa$, while its position remains unchanged. This might be expected as the departure of $\kappa_{I}$ from $\kappa_c$ increases the magnitude of the quench, and hence increases the final energy density, thereby blunting the signature of the quantum phase transition. This is consistent with the results in Sec.~\ref{sec:kappa_I}, where we discuss the effect that the increase in the magnitude of the quench has in observables after thermalization.

\section{Results after thermalization}\label{sec:thermalization}

Because of the linear growth of the entanglement entropy seen in Fig.~\ref{Others}, the iTEBD technique only allows one to study dynamics at short and intermediate times. To explore the fate of observables after thermalization, we use a numerical linked cluster expansion (NLCE)~\cite{rigol2006numerical, *rigol2007numerical}. We broaden the class of initial states to explore how initial nonzero temperatures modify the behavior of observables after thermalization.

Here we consider more general quenches within the ANNNI Hamiltonian involving initial states $\hat{\rho}_I$ that are thermal equilibrium states of the initial Hamiltonian $\hat{H}(\kappa_I)$. For an initial temperature $T_I$, $\hat{\rho}_I$ has the form
\begin{eqnarray}\label{eq:rhoi}
\hat{\rho}_I=\dfrac{e^{-\hat{H}(\kappa_I)/T_I}}{\text{Tr}[e^{-\hat{H}(\kappa_I)/T_I}]}.
\end{eqnarray} 
When $T_I=0$, $\hat{\rho}_I$ is the ground state of $\hat{H}(\kappa_I)$. As in the previous section, we quench $\kappa_I\rightarrow\kappa$, while $\Gamma$ is kept unchanged ($\Gamma=0.2$). In Secs.~\ref{sec:observables} and~\ref{sec:temperature} we fix $\kappa_I=0$. In Sec.~\ref{sec:kappa_I}, we explore what changes when $\kappa_I$ is varied within the ferromagnetic phase ($\kappa_I<\kappa_c\approx0.41$).

Since the energy after the quench is the only conserved quantity, at sufficiently long times in the thermodynamic limit, observables are expected to be described by a Gibbs ensemble~\cite{dalessio_kafri_16}
\begin{eqnarray}\label{eq:GE}
\hat{\rho}_{\text{GE}}(\kappa)=\dfrac{e^{-\hat{H}(\kappa)/T(\kappa)}}{\text{Tr}[e^{-\hat{H}(\kappa)/T(\kappa)}]},
\end{eqnarray} 
with a temperature $T(\kappa)>0$ (which is nonzero even when $T_I=0$) determined by the energy $E(\kappa)$ set by the initial state $\hat{\rho}_I$, as dictated by:
\begin{eqnarray}
\text{Tr}[\hat{\rho}_{\text{GE}}(\kappa) \hat{H}(\kappa)]=\text{Tr}[\hat{\rho}_I \hat{H}(\kappa)]\label{E_match}.
\end{eqnarray} 

We use the numerical linked cluster expansion (NLCE) technique introduced in Refs.~\cite{rigol2006numerical, *rigol2007numerical} to study the thermal expectation values of observables in the thermodynamic limit (see Appendix~\ref{Apndx:comp} for details). All the NLCE results for the ANNNI chain are obtained using 15 orders of the maximally connected cluster expansion introduced in Refs.~\cite{rigol_14a,*rigol_16}. To gauge how well the series has converged, we estimate the convergence error for an observable by computing the relative difference between the last two orders (14 and 15) of the NLCE~\cite{rigol_14a,*rigol_16}. We only report results whose convergence error for the energy is less than $10^{-5}$. $T(\kappa)$ is obtained by numerically matching the energies in the left and right side of Eq.~\eqref{E_match}. Both energies are evaluated using NLCE to 15 orders, and $T(\kappa)$ is computed by enforcing that their relative difference is less than $10^{-11}$ (see Ref.~\cite{rigol_14a,*rigol_16}). For observables other than the energy, we only report results whose convergence errors are less than $5\times 10^{-5}$ (except for the entropy, for which we set the cut off to be $7\times 10^{-5}$). Those errors are small enough to be unimportant for the discussions that follow. 

\subsection{Observables}\label{sec:observables}

As mentioned before, in the thermodynamic limit at sufficiently long times after the quench, thermalization is expected to occur in the nonintegrable systems considered here~\cite{dalessio_kafri_16}. Next we study the expected thermal equilibrium results that observables $\hat O$ reach after equilibration following the quench. 

In the space of all possible thermal equilibrium ensembles parameterized by the coordinates $(T,\kappa)$, the initial state $\hat{\rho}_I$ sets a trajectory $T(\kappa)$ determined by Eq.~\eqref{E_match}. One can then write
\begin{eqnarray}
\dfrac{dO}{d\kappa}&=& \dfrac{dT}{d\kappa}\left(\dfrac{\partial O}{\partial T}\right)_{\kappa}+\left(\dfrac{\partial O}{\partial \kappa}\right)_{T}\label{O_partial}.
\end{eqnarray}
Since $O(T,\kappa)$ is an analytic function whenever $T>0$, and since $dT/d\kappa$ is expected to be a smooth function of $\kappa$ (we discuss this in Sec.~\ref{sec:temperature}), then $dO/d\kappa$ must be a smooth function of $\kappa$ after equilibration following the quench. Still, for observables that are indicators of the quantum phase transition in nonintegrable systems (e.g., order parameters and related observables), $(\partial O/\partial T)_{\kappa}$ and $(\partial O/\partial \kappa)_{T}$ can be large if $T$ is low when $\kappa$ is close to $\kappa_c$ (we show the latter to be the case for our quenches in Sec.~\ref{sec:temperature}). This means that, even in thermal equilibrium, it is possible to have prominent (but smooth) features in $dO/d\kappa$ as observed at intermediate times in the previous section. In integrable systems, in which all possible states after equilibration are described by generalized Gibbs ensembles that are parameterized by extensive numbers of quantities~\cite{rigol_dunjko_07, vidmar_rigol_16}, nonanalytic behavior is possible and has in fact been observed in Refs.~\cite{SciRep_BDD, Topological_QPTSgn_RMD}.

\begin{figure}[!t]
\includegraphics[width=0.99\linewidth]{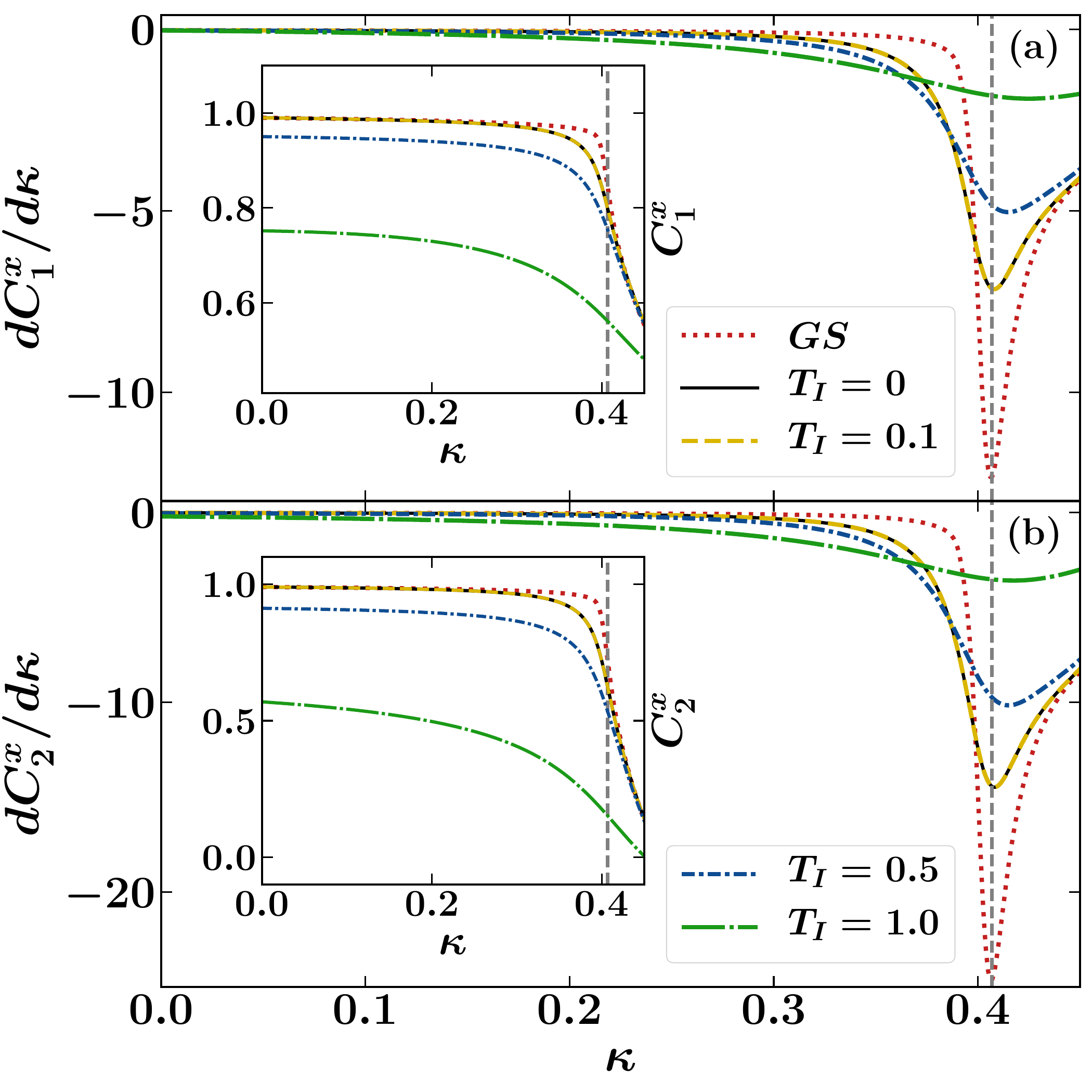}
  \caption{The nearest (next nearest) neighbor longitudinal spin-spin correlation per site $C_{1}^{x}$ ($C_{2}^{x}$), see Eq.~\eqref{Cx}, evaluated in thermal equilibrium using NLCE following quenches $\kappa_{I}=0\rightarrow\kappa$, with $T_I=0,\,0.1,\,0.5,$ and 1.0. We also show $C_{1}^{x}$ and $C_{2}^{x}$ in the ground state of $\hat{H}(\kappa)$ (dotted lines) computed using iTEBD. The main panels in (a) and (b) show $dC_{1}/d\kappa$ and $dC_{2}/d\kappa$, respectively, while the corresponding insets show $C_{1}^{x}$ and $C_{2}^{x}$. The vertical dashed lines mark the critical $\kappa_c\approx0.41$.}
\label{fig:corr_vs_kappa} 
\end{figure}

In Fig.~\ref{fig:corr_vs_kappa}, we show the thermal equilibrium results obtained for the nearest $C_{1}^{x}$ and next nearest $C_{2}^{x}$ neighbor longitudinal spin correlations per site [see Eq.~\eqref{Cx}] as functions of $\kappa$ after the quench, as well as their expectation values in the ground state of $\hat{H}(\kappa)$ computed with iTEBD. The main panels show $dC_{1(2)}/d\kappa$, while the insets show $C_{1 (2)}(\kappa)$, for various initial temperatures $T_I$ and in the ground state (dotted lines, computed with iTEBD). In the ground state, $C_{1}^{x}$ and $C_{2}^{x}$ are nearly one in the ferromagnetic phase and exhibit a rapid decrease when crossing the quantum phase transition (prominent minima can be seen in $dC_{1(2)}/d\kappa$ at $\kappa_c$), i.e., they serve as indicators of the quantum phase transition. (They also serve as indicators of the ferromagnetic to paramagnetic quantum phase transition in the integrable transverse field Ising model, see Appendix~\ref{Apndx:Ising}.) This zero-temperature behavior is the precursor of the behavior of $C_{1}^{x}$ and $C_{2}^{x}$ observed in the insets for low initial $T_I$, which, in turn, produces the prominent minima in $dC_{1(2)}/d\kappa$ near $\kappa_c$ observed in the main panels. Figure~\ref{fig:corr_vs_kappa} shows that the position of the minima drift away from $\kappa_c$, and they become shallower, with increasing $T_I$. Note that the results for $T_I=0$ and $T_I=0.1$ overlap in the plots. 

\begin{figure}[!t]
\includegraphics[width=0.99\linewidth]{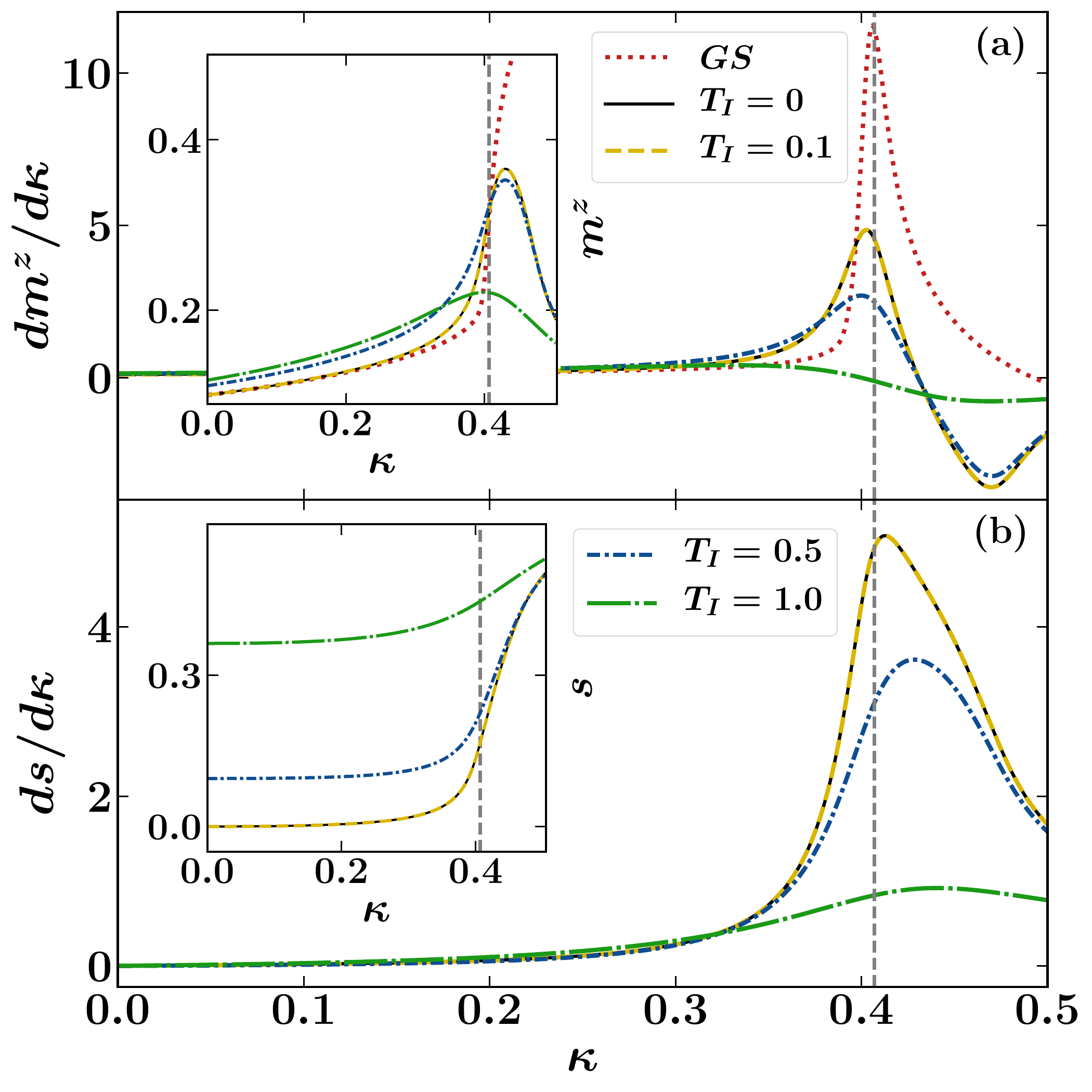}
  \caption{The transverse spin magnetization ($m^z$) and the von-Neumann entropy ($s$), per site (see text), evaluated using NLCE in thermal equilibrium following quenches $\kappa_{I}=0\rightarrow\kappa$ with $T_I=0,\,0.1,\,0.5,$ and 1.0. The main panels in (a) and (b) show $dm^{z}/d\kappa$ and $ds/d\kappa$, respectively, while the corresponding insets show $m^z$ and $s$. In (a), we also show results for $m^z$ in the ground state of $\hat{H}(\kappa)$ (dotted lines) computed using iTEBD. The vertical dashed line marks the critical $\kappa_c\approx0.41$.}
\label{fig:obs_vs_kappa} 
\end{figure}

Qualitatively similar results were obtained for other local observables, such as the transverse magnetization per site $m^z=\sum_i \sigma_{i}^{z}/L$, shown in Fig.~\ref{fig:obs_vs_kappa}(a), and for the (von-Neumann) entropy per site $s=-\text{tr}(\hat{\rho}_\text{GE}\ln\hat{\rho}_\text{GE})/L$ of the thermal state $\hat{\rho}_\text{GE}(\kappa)$, shown in Fig.~\ref{fig:obs_vs_kappa}(b). In the ground state, $m^z$ increases rapidly when transitioning from the ferromagnetic to the paramagnetic phase, as shown in Fig.~\ref{fig:obs_vs_kappa}(a) (dotted lines, computed with iTEBD). Hence, $m^z$ serves as an indicator of the quantum phase transition, and its behavior at zero temperature is the reason there are prominent maxima in $dm^{z}/d\kappa$ near $\kappa_c$ for quenches at low $T_I$. (See Appendix~\ref{Apndx:Ising} for ground-state results of $m^z$ across the ferromagnetic to paramagnetic quantum phase transition in the integrable transverse field Ising model.) The entropy, on the other hand, is strictly zero at zero temperature, i.e., it does not change at the phase transition [the entanglement entropy does change, as shown in Fig.~\ref{Others}(b)]. However, as we show in Sec.~\ref{sec:temperature}, when $T_I$ is low, the temperature after quench increases rapidly when $\kappa$ crosses $\kappa_c$ and this produces the rapid increase of $s$ seen in Fig.~\ref{fig:obs_vs_kappa}(b).

\subsection{Temperature}
\label{sec:temperature}
Let us now show that $dT/d\kappa$ is a smooth function of $\kappa$. The ANNNI Hamiltonian can be written as $\hat{H}(\kappa)=\hat{H}_0+\kappa\hat{V}$, so that keeping the initial state $\hat{\rho}_I$ fixed and changing $\kappa$ after the quench results in $E(\kappa)$ being a linear function of $\kappa$
\begin{eqnarray}
E(\kappa)=\left(\text{Tr}[\hat{\rho}_I \hat{H}_0]\right)+\kappa\left(\text{Tr}[\hat{\rho}_I \hat{V}]\right)\label{linearE},
\end{eqnarray}
with a slope $A\equiv dE(\kappa)/d\kappa=\text{Tr}[\hat{\rho}_I \hat{V}]$. 

As in the previous section for $dO/d\kappa$, for the energy one can write
\begin{eqnarray}
\dfrac{dE}{d\kappa}&=& \dfrac{dT}{d\kappa}\left(\dfrac{\partial E}{\partial T}\right)_{\kappa}+\left(\dfrac{\partial E}{\partial \kappa}\right)_{T}\label{E_partial},
\end{eqnarray}
where $\left(\partial E/\partial T\right)_{\kappa}=C_{\kappa}(T)$ is the specific heat. Combining Eqs.~\eqref{linearE} and~\eqref{E_partial}, we have that 
\begin{eqnarray}
\dfrac{dT(\kappa)}{d\kappa}=\dfrac{A-\left(\dfrac{\partial E}{\partial \kappa}\right)_{T(\kappa)}}{C_\kappa[T(\kappa)]}\label{dtdk(TK)}.
\end{eqnarray}
All functions in the r.h.s.~of Eq.~\eqref{dtdk(TK)} are smooth, and $C_\kappa[T(\kappa)]>0$, because $T(\kappa)>0$ after the quench. This shows that $T(\kappa)$ is also a smooth function. Next, we use numerical calculations to explore whether quenches $\kappa_I\rightarrow\kappa$ spanning across $\kappa_c$ produce temperatures $T(\kappa)$ with signatures of the quantum phase transition, as shown to be the case in Sec.~\ref{sec:observables} for local observables.

\begin{figure}[!t]
\includegraphics[width=0.99\linewidth]{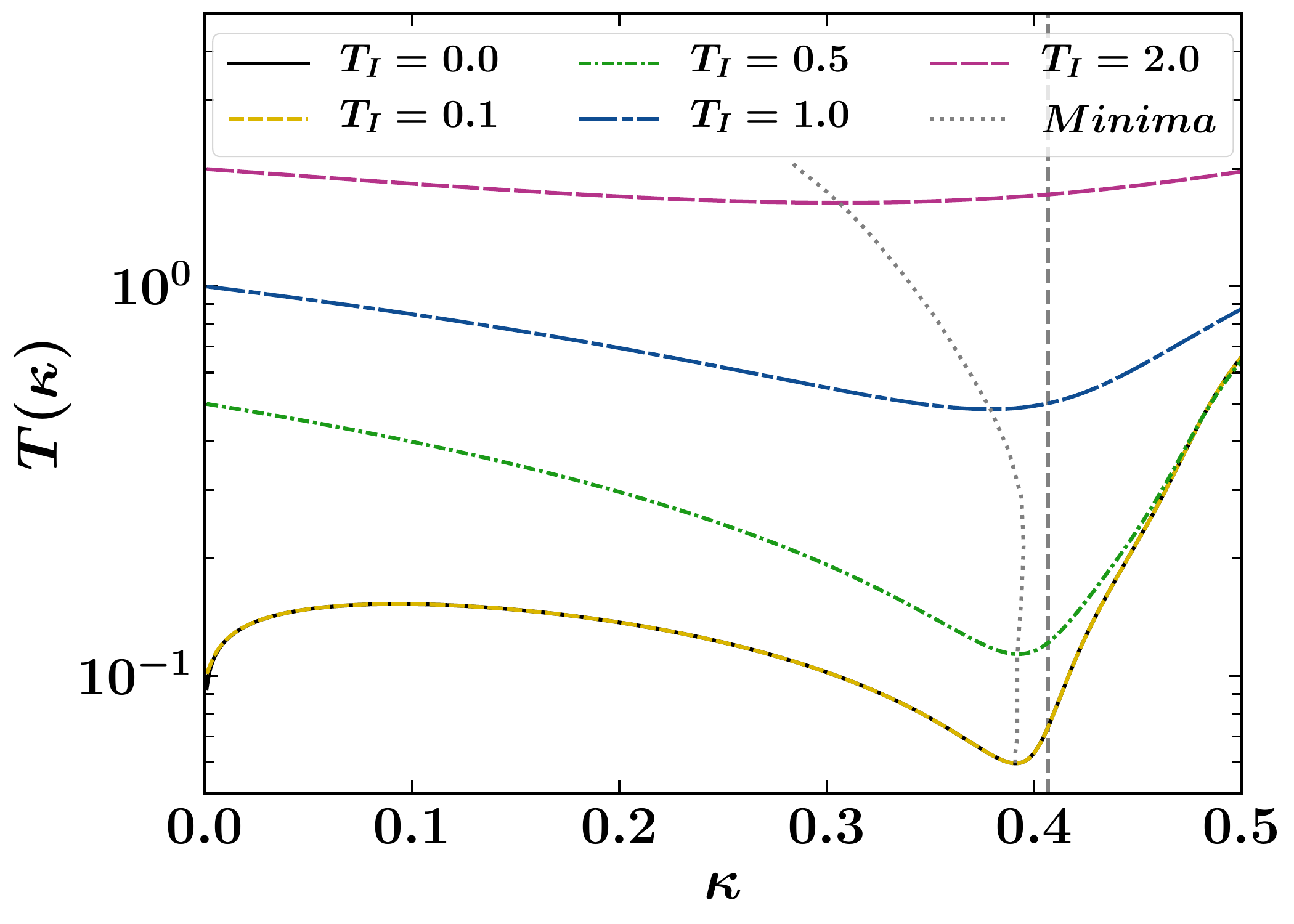}
  \caption{NLCE results for the temperature of the Gibbs ensemble describing observables after equilibration, following quantum quenches $\kappa_I=0\rightarrow\kappa$ within the ANNNI Hamiltonian, for initial thermal states at \marcos{temperatures $T_I=0.0$, 0.1, 0.5, 1.0, and 2.0. For those initial temperatures, minima in $T(\kappa)$ occur at $\kappa_m\approx0.391$, 0.391, 0.392, 0.379, and 0.306, respectively}. 
  The locus of minima [$\kappa_m,T(\kappa_m)$] for a large number of initial temperatures $T_I$ is also shown. The vertical dashed lines mark the critical $\kappa_c\approx0.41$.}
\label{fig:T(k)}
\end{figure}

Figure~\ref{fig:T(k)} shows $T(\kappa)$ for quenches with $\kappa_I=0\rightarrow\kappa$ for various initial temperatures $T_I$, including the ground state of $\hat{H}(\kappa_I)$. For very low initial temperatures $T_I\lesssim 0.1$, the temperatures $T(\kappa)$ after the quench are essentially indistinguishable from those for $T_I=0$. This explains why all the results reported in Sec.~\ref{sec:observables} are indistinguishable for $T_I=0$ and $T_I=0.1$. For those very low $T_I$, the temperatures $T(\kappa)$ exhibit a low-temperature minimum in the vicinity of $\kappa_c$ \marcos{(at $\kappa_m\approx0.39$, for which $T(\kappa_m)\approx 0.06$)}. 
At $T_I=0.5$, a temperature at which $T(\kappa)$ after the quench departs from the $T_I=0$ result, a minimum in $T(\kappa)$ still remains visible close to $\kappa_c$. The locus of minima in $T(\kappa)$, shown as a dotted line for a large number of $T_I$, makes apparent that the minima remain close to $\kappa_c$ as long as $T_I$ remains low ($T_I\lesssim1.0$). At higher initial temperatures, the minima depart from $\kappa_c$ indicating that the information about $\kappa_c$ is washed out.

Overall it is remarkable that, due to the presence of the phase transition (and the corresponding closing of the gap above the ground state), when quenching to the same (ordered) side of the critical point, the effective temperature decreases as the size of the quench increases and $\kappa$ approaches the critical point. This trend sharply reverses as $\kappa$ crosses the critical point. Examining Eq.~\eqref{dtdk(TK)} in the context of our numerical results allows us to understand why a minimum develops near $\kappa_c$ at very low ($T_I\lesssim 0.1$) and low ($T_I\lesssim 1.0$) initial temperatures. At the minimum, we have that
\begin{eqnarray}\label{eq:min}
a=\left(\dfrac{\partial e}{\partial \kappa}\right)_{T},
\end{eqnarray}
where we defined the intensive counterparts of the extensive quantities in Eq.~\eqref{dtdk(TK)} as $a=A/L$ and $e=E/L$. 

\begin{figure}[!t]
\includegraphics[width=0.98\linewidth]{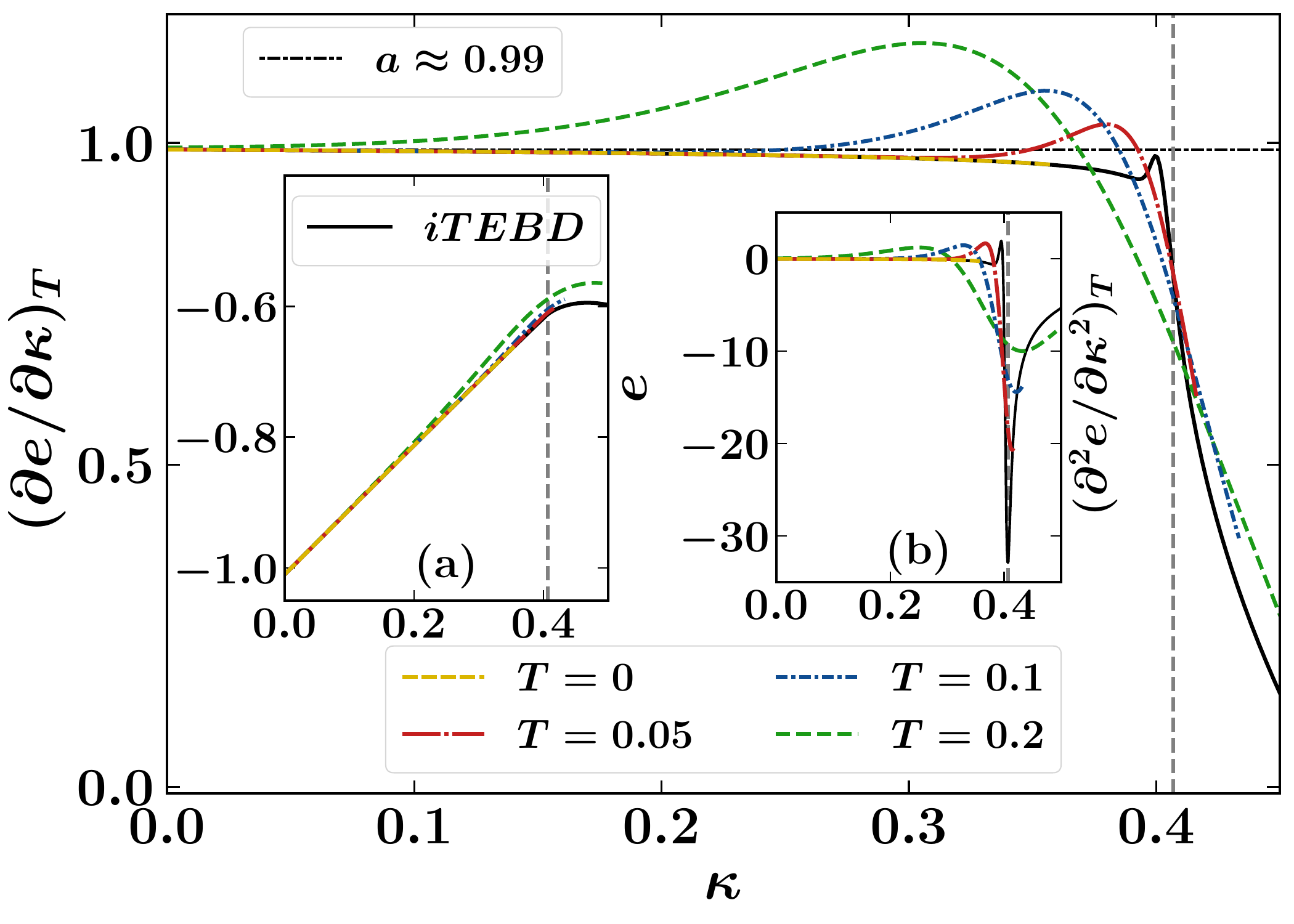}
  \caption{Thermal equilibrium energy per site $e$ and its derivatives at different $T$. The main panel shows $(\partial e/\partial \kappa)_T$, inset (a) shows $e$ versus $\kappa$ at constant temperature, and inset (b) shows $(\partial^{2} e/\partial \kappa^{2})_T$. Results are shown for different values of $T$ (the main panel and the insets share the legend). The solid (black) curve ($T=0$) shows the iTEBD results for the ground state, while the other curves show NLCE results. Also depicted in the main panel is $a=\langle \sigma_{i}^{x}\sigma_{i+2}^{x}\rangle_{\hat{\rho}_I}/L\approx 0.99$ for $T_I\lesssim 0.1$. The vertical dashed lines mark the critical $\kappa_c\approx0.41$.}
\label{fig:E_vs_kappa}
\end{figure}

The main panel in Fig.~\ref{fig:E_vs_kappa} shows $(\partial e/\partial \kappa)_T$ vs $\kappa$ at different temperatures [inset Fig.~\ref{fig:E_vs_kappa}(a) shows $e$ vs $\kappa$ at the same temperatures]. For $T=0$, we also show iTEBD results (the NLCE results do not converge close to $\kappa=\kappa_c$). Notice that, in the region in which the NLCE results converge to the precision mentioned in the introduction of this section, they are indistinguishable from the iTEBD ones. The iTEBD results for $(\partial e/\partial \kappa)_{T=0}$ exhibit a rapid decrease about $\kappa_c$ [resulting in a singularity in $(\partial^2 e/\partial \kappa^2)_{T=0}$ at $\kappa_c$, as shown in inset Fig.~\ref{fig:E_vs_kappa}(b)], reflecting the nonanalytic behavior of the energy at the (second order) quantum phase transition. That rapid decrease leaves its signature in the low-temperature behavior of $(\partial e/\partial \kappa)_{T>0}$, and this is what makes possible for Eq.~\eqref{eq:min} to be satisfied close to $\kappa_c$ for low initial temperatures. 

In Fig.~\ref{fig:E_vs_kappa}, $a=\langle \sigma^{x}_{i} \sigma^{x}_{i+2} \rangle_{\hat{\rho}_I}/L$ is shown as a horizontal line for $T_I\lesssim 0.1$. For those very low initial temperatures, $a$ is very close to 1 ($a\approx 0.99$) since $\kappa_I=0$ is deep in the ferromagnetic phase, and Fig.~\ref{fig:E_vs_kappa} shows that the condition $(\partial e/\partial \kappa)_{T}=a$ is satisfied \marcos{at $\kappa=0.39$ for $T=0.05$ and at $\kappa=0.37$ for $T=0.1$. Those two temperatures approximately bound the range of effective temperatures after the quench for $\kappa$ close to $\kappa_c$ when $T_I\lesssim 0.1$, see Fig.~\ref{fig:T(k)}}. 
This explains why the minimum in $T(\kappa)$ vs $\kappa$ occurs very close to $\kappa_c$ for $T_I\lesssim 0.1$. Increasing the initial temperature beyond $T_I=0.1$ increases $T$ but also reduces the value of $a$. This results in the minimum remaining close (and actually slightly approaching) $\kappa_c$ in Fig.~\ref{fig:T(k)} when $T_I$ departs from 0.1 but still remains low ($T_I\lesssim1.0$). Since the slope of $(\partial e/\partial \kappa)_{T}$ at the crossing point near $\kappa_c$ is negative, it follows from Eq.~\eqref{dtdk(TK)} that the extremum in $T(\kappa)$ near $\kappa_c$ is a minimum.

\subsection{Changing $\kappa_I$}\label{sec:kappa_I}

\begin{figure}[!t]
\includegraphics[width=0.99\linewidth]{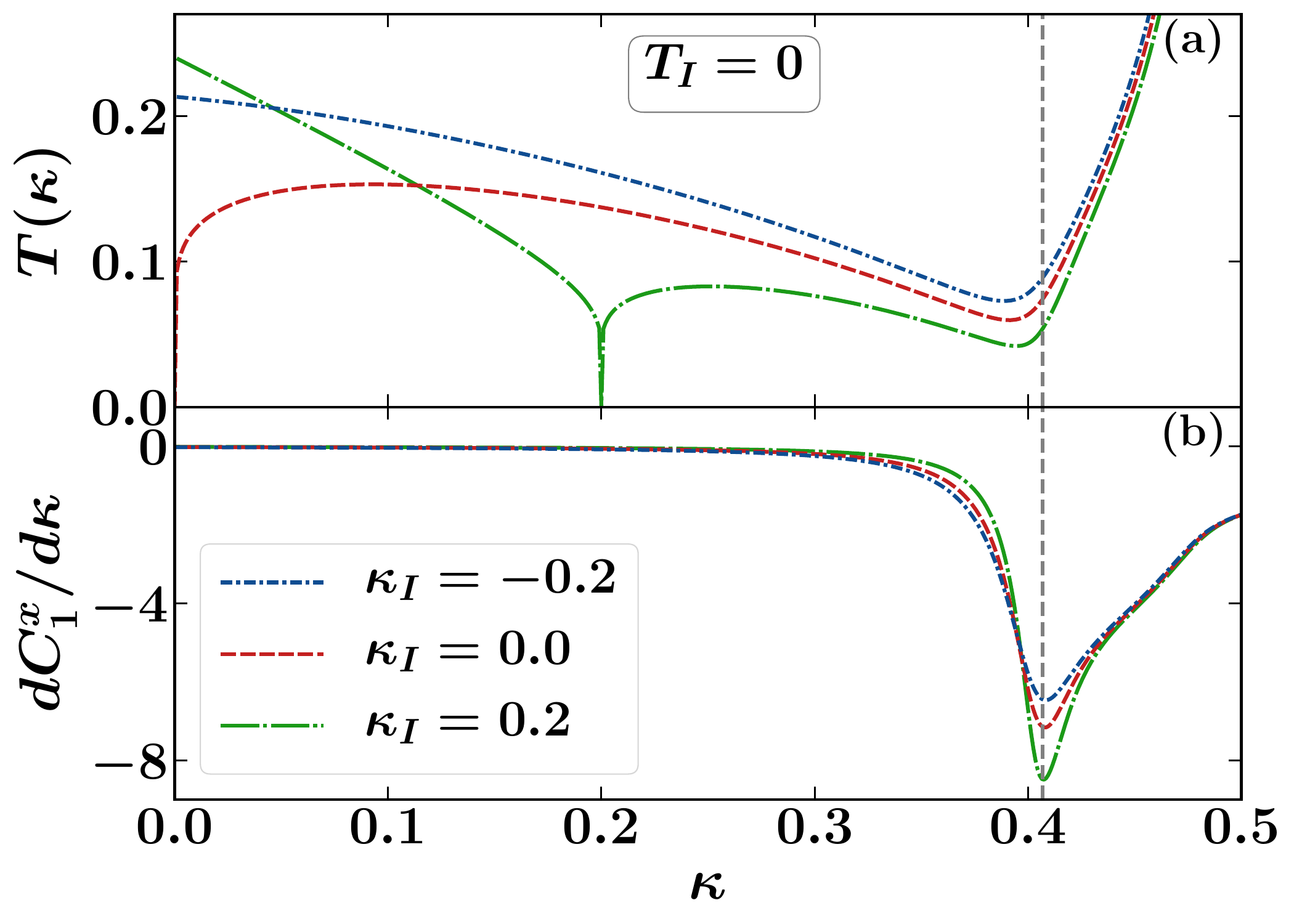}
  \caption{(a) Equilibrium temperature $T(\kappa)$ and (b) $dC_{1}^{x}/d\kappa$ in thermal equilibrium, after quenches $\kappa_I\rightarrow \kappa$ from initial ground states of $\hat{H}(\kappa_I)$ for three different values of $\kappa_I$. The vertical dashed lines mark the critical $\kappa_c\approx0.41$.} 
\label{fig:obs_vs_kappa_ini}
\end{figure}

Motivated by the results discussed in Sec.~\ref{sec:timekappa_I}, we explore next what happens to the thermal equilibrium results after equilibration when one changes $\kappa_I$ within the ferromagnetic regime, keeping $T_I=0$ fixed. In Fig.~\ref{fig:obs_vs_kappa_ini}(a), we show $T(\kappa)$ vs $\kappa$ for $\kappa_I=-0.2$, 0, and $0.2$. As expected from the fact that the initial state remains a nearly perfect ferromagnet, the minima in $T(\kappa)$ close $\kappa_c$ are robust to the choice of initial $\kappa_I$. However the minimum value of $T(\kappa)$ attained decreases as $\kappa_I$ approaches $\kappa_c$. As a result, the signature of the presence of a quantum critical point in observables after thermalization becomes sharper as $\kappa_I\rightarrow \kappa_c$. This is apparent in Fig.~\ref{fig:obs_vs_kappa_ini}(b) in which we plot $dC_{1}^{x}/d\kappa$. 

Note that in Fig.~\ref{fig:obs_vs_kappa_ini}(a) there is a singularity in $T(\kappa)$ at $\kappa=0.2$ for $\kappa_I=0.2$, as well as at $\kappa=0$ for $\kappa_I=0$. These are trivial consequences of performing no quench, which means that the system remains in the ground state. The fact that $|dT/d\kappa|\rightarrow\infty$ at those points follows from Eq.~\eqref{dtdk(TK)} due to specific heat $C_\kappa(T\rightarrow0)\rightarrow0$ in the denominator. These singularities have no consequence in the expectation values of observables.

\section{Phase Diagram}\label{sec:phasediagram}

Here we combine results obtained for $C_{1}^{x}$ at intermediate times after the quench (from iTEBD calculations), and after equilibration (from NLCE calculations), to identify, in the $(\kappa,\Gamma)$ plane, the phase boundary separating the ferromagnetic and paramagnetic phases in the ground state. We estimate $\kappa_c$ by carrying out quenches $\kappa_I = 0\rightarrow\kappa$ for different values of $\Gamma$ ($\Gamma$ is not changed during the quench). Qualitatively similar results were obtained for other local observables such as $C_{2}^{x}$ and $m^z$ and are not reported here. 

\begin{figure}[!t]
\includegraphics[width=0.99\linewidth]{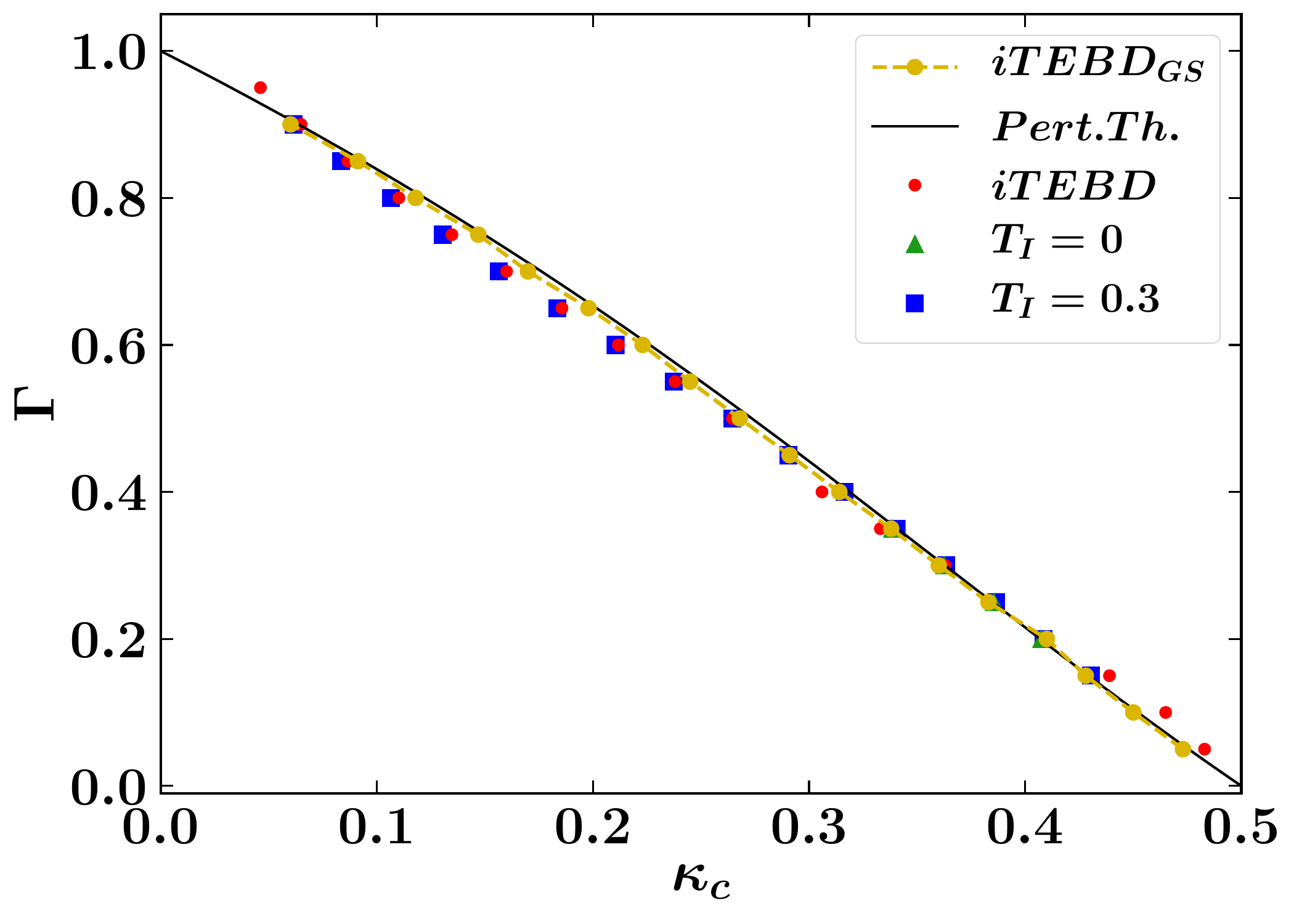}
  \caption{Phase boundary for the ground-state quantum phase transition separating the ferromagnetic and paramagnetic phases in the $(\kappa, \Gamma)$ plane. Unbiased results for the boundary were obtained using ground-state iTEBD [iTEBD$_\text{GS}$ in the legend, obtained locating the singularity in $(\partial^2 e/\partial \kappa^2)_{T=0}$] and are closely followed by the predictions of second order perturbation theory (continuous line). The phase boundary is well described by $\kappa_c$ estimated from the extrema of $dC_{1}^{x}/d\kappa$ obtained in finite-time iTEBD calculations after the quench (for $t = 25$) and in the (expected) long-time thermal results obtained using NLCE. In all quenches $\kappa_I=0\rightarrow\kappa$, $\Gamma$ is not changed during the quench, and we show results for $T_I=0$ (iTEBD and NLCE), and for $T_I=0.3$ (NLCE).}
\label{fig:phase_obs}
\end{figure}

In the main panel of Fig.~\ref{fig:phase_obs}, we show $\kappa_c$ extracted from the extrema of $dC_{1}^{x}/d\kappa$ obtained using iTEBD results at $t = 25$ after quenches starting from the ground state, and NLCE thermal equilibrium results after quenches starting from the ground state ($T_I=0$) and from an initial temperature $T_I=0.3$. As $\Gamma$ increases, the NLCE convergence errors are higher for quenches starting from the ground state. This occurs because the critical point gets closer to $\kappa_I=0$ and the effective temperature after the quench becomes too small (see Fig.~\ref{fig:obs_vs_kappa_ini} and related discussion). This is the reason no NLCE points are reported for quenches with $\Gamma\ge0.4$ and $T_I=0$. On the other side of the phase diagram, when $\Gamma$ is small, the quenches in $\kappa$ result in fewer excitations ($\Gamma\rightarrow0$ becomes the classical Ising chain) thereby bringing the thermal equilibrium ensemble about $\kappa_c$ close to the ground-state critical point. This also affects the NLCE convergence, resulting in no NLCE data points for $\Gamma\lesssim 0.2$. The results in Fig.~\ref{fig:phase_obs} show that both the intermediate-time and (expected) long-time extrema follow very closely the phase boundary calculated using iTEBD for the ground state [locating the singularity in $(\partial^2 e/\partial \kappa^2)_{T=0}$], which is well described by the second order perturbation theory results.

\section{Experimental tests}\label{sec:experiments}
It is a central aspect of this work that the reported signatures of the quantum phase transitions in the ANNNI model are accessible in state-of-the-art quantum simulator platforms with Rydberg atoms. The ANNNI Hamiltonian~(\ref{H_ANNNI}) can be straightforwardly realized using using Rydberg dressing in ultracold atoms in optical lattices~\cite{2016Zeiher, 2017Zeiher}. \blue{Rydberg-dressed atoms exhibit a soft-core interaction potential $J_{i,j} = J_0/[1+(R_{ij}/R_c)^6]$, which is approximately constant below a threshold distance $R_c$ between two atoms and decays quickly beyond the threshold $R_c$ (in a $R_{ij}^{-6}$ fashion as a function of distance $R_{ij} = a |i-j|$, where $a$ is the lattice spacing between the involved spins)~\cite{2016Zeiher}. Realizing approximately the ANNNI model with such a soft-core interaction potential requires to choose the tunable threshold $R_c$ such that $(R_{i,i+3}/R_c)^6 \gg (R_{i,i+2}/R_c)^6, (R_{i,i+1}/R_c)^6$ so that $J_{i,i+3} \sim J_0 (R_{i,i+2}/R_c)^{-6} \ll J_{i+1}, J_{i+2}$. In such a regime only nearest and next-nearest neighbor couplings have to be taken into account, while further distant ones can be neglected. The relative strength of nearest and next-nearest neighbor interactions, quantified by $\kappa=J_{i,i+2}/J_{i,i+1} = [1+(a/R_c)^6]/[1+(2a/R_c)^6]$ in Eq.~(\ref{H_ANNNI}), can also be varied by tuning $R_c$ relative to the lattice spacing $a$ with the only limitation that $\kappa<1$. As the targeted quantum critical point $\kappa_c\approx 0.41 < 1$, the reported signatures therefore lie within the tunability of the couplings. Let us note that the interaction in the experiment would be of antiferromagnetic nature and not directly of the type required in Eq.~(\ref{H_ANNNI}). However, by performing a rotation $\sigma_l^x \to -\sigma_l^x$ on every other lattice site, e.g., even ones, the Hamiltonian in Eq.~(\ref{H_ANNNI}) maps onto a purely antiferromagnetic spin model and therefore to the one which can be realized experimentally. Furthermore, transverse fields can be straightforwardly generated, implying that the full Hamiltonian can be modeled with high accuracy.}

\blue{It remains to clarify whether also the dynamics of this system can be accessed in the desired regimes, which we now answer in the affirmative. First, Rydberg-dressed atom systems with a large number of spins ($L\approx 200$) were already created in Ref.~\cite{2016Zeiher}. The trapping potential for the ultracold quantum gas only has a minor impact when considering Rydberg dressing, it affects the preparation of the initial condition by limiting the maximal number of spins which can be controllably initialized~\cite{2016Zeiher}. Specifically, the fully polarized initial condition we are considering in our work can be prepared with high fidelity as demonstrated in Ref.~\cite{2017Zeiher}.} Hence, the main point that remains to be addressed is the coherence time, i.e., whether it is possible to identify the proposed signatures before decoherence sets in. In a recent experiment with Rydberg-dressed atoms time scales $Jt \gtrsim 10$ were achieved, where $J$ denotes the strength of the nearest-neighbor couplings. Consequently, the time scales discussed in Sec.~\ref{sec:dynamics} are in the experimentally accessible regime. We note that also the desired spin-spin correlation functions in Eq.~(\ref{Cx}) can be measured in the aforementioned experimental systems~\cite{2017Zeiher}.

\section{Topological Transitions}\label{sec:topological}

A final question we address next is how generally one can use the previously introduced protocol to locate quantum phase transitions in one-dimensional models. Given the results obtained and insights gained within the ANNNI chain (notice that in Fig.~\ref{fig:phase_obs} we report results for an entire phase boundary), we expect this protocol to be widely applicable to one-dimensional models with traditional quantum phase transitions. A different question is whether such signatures in local quantities can be used to locate topological quantum phase transitions, as shown for noninteracting models in Ref.~\cite{Topological_QPTSgn_RMD} (non-local
quantities can, of course, retain such information in
the noninteracting case -- see, e.g., ~\cite{Haitao_1,Dag_1}). In what follows, we report results from a preliminary exploration of dynamics after quantum quenches about topological transitions in two quantum-chaotic models.

First, we explore the quantum phase transition from the N\'{e}el to the symmetry protected topological ``Haldane" phase in the spin-1 anisotropic (XXZ) Heisenberg chain model. The Hamiltonian for this model reads
\be\label{XXZ_Ham}
\hat H_\text{XXZ} = \sum_{i}^{L}\left(\hat S_{i}^{x}\hat S_{i+1}^{x} + \hat S_{i}^{y}\hat S_{i+1}^{y} + \Delta \hat S_{i}^{z} \hat S_{i+1}^{z} \right),
\ee
where $\hat S^{x,y,z}_{i}$ denote the $x,\,y$ and $z$ components of the spin-1 operator at site $i$. Four different phases occur in this model when one changes the anisotropy parameter $\Delta$ (see, e.g., Refs.~\cite{Chen_XXZD,Zhou_XXZ} and references therein). Here we focus on the transition that occurs upon decreasing $\Delta$ from $\Delta > 1$, a limit in which $\hat H_\text{XXZ}$ reduces to the spin-1 Ising antiferromagnet. With decreasing $\Delta$, the ground state of $\hat H_\text{XXZ}$ undergoes a quantum phase transition from the antiferromagnet to the Haldane phase at $\Delta_{c} \approx 1.183$. The Haldane phase is a topological phase, protected by any one of the following three global symmetries: $D_{2}$ spin rotation, time-reversal, and bond centered inversion~\cite{Frank_Haldane}. This transition is of second order, and belongs to the 2D Ising universality class~\cite{XXZ_Takahashi, XXZ_Nomura}. 

\begin{figure}[!t]
\includegraphics[width=0.99\linewidth]{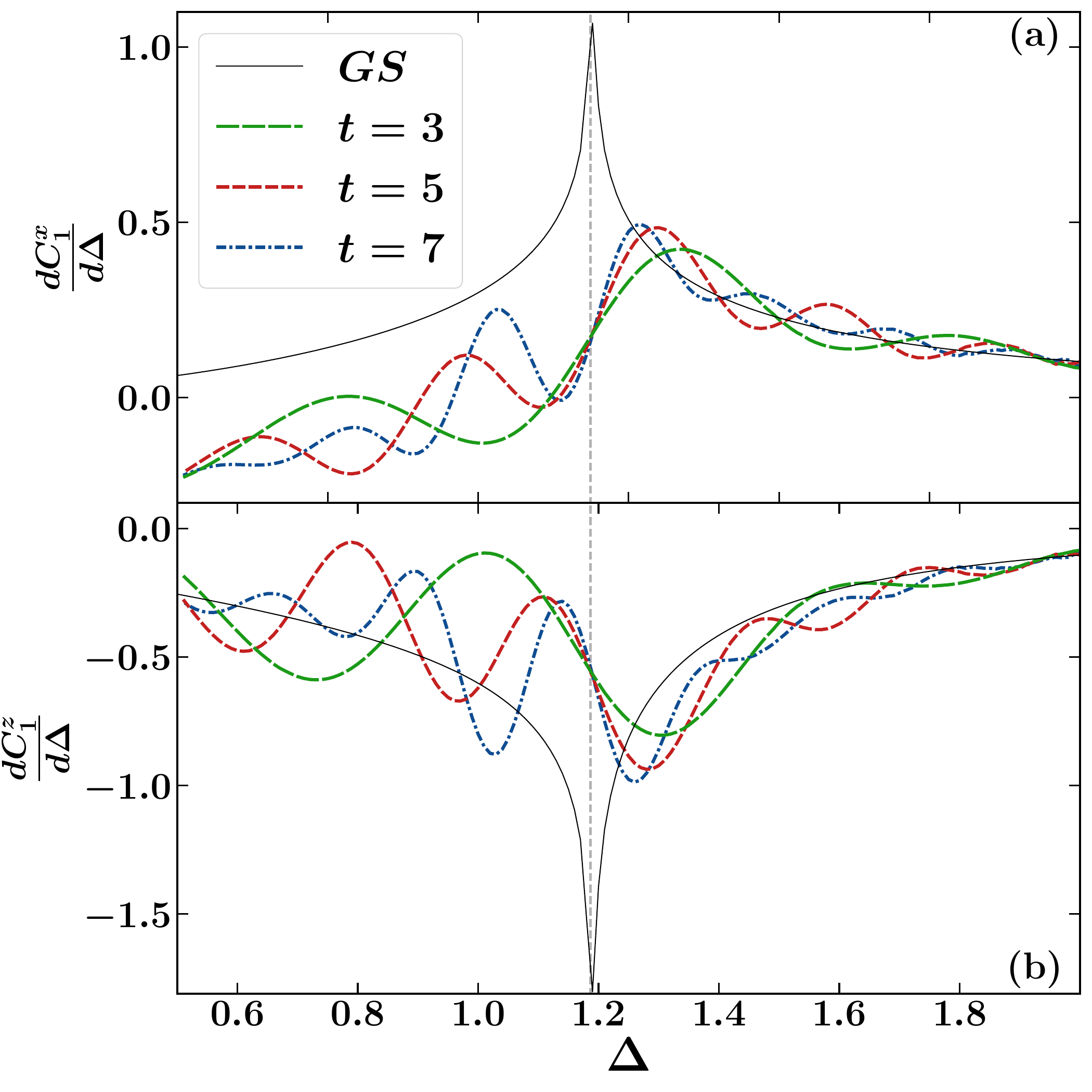}
  \caption{Signatures of the N\'{e}el to Haldane quantum phase transition in the anisotropic XXZ chain. Derivatives with respect to $\Delta$ of (a) $C_{1}^{x}$ [see Eq.~\eqref{Cx}] and (b) $C_{1}^{z}$ [see Eq.~\eqref{Cz}] at different fixed times after the quench, and of the results in the ground state (black solid line). The gray vertical line shows the critical $\Delta_c \approx 1.183$}
\label{Topological}
\end{figure}

In Fig.~\ref{Topological}, we show ground state results for $dC_1^x/d\Delta$, where
\beq\label{eq:Cx}
C^{x}_{1} = \frac{1}{L}\sum_{i=1}^{L}\langle\hat S_{i}^{x} \hat S_{i+1}^{x}\rangle,
\eeq
and for $dC_1^z/d\Delta$, where
\beq\label{Cz}
C^{z}_{1} = \frac{1}{L}\sum_{i=1}^{L}\langle\hat S_{i}^{z} \hat S_{i+1}^{z}\rangle,
\eeq
plotted as functions of $\Delta$. As for the local observables shown in Fig.~\ref{fig:corr_vs_kappa} for the ANNNI model, $dC_1^x/d\Delta$ in Fig.~\ref{Topological}(a) [$dC_1^z/d\Delta$ in Fig.~\ref{Topological}(b)] exhibits a sharp maximum (minimum) at the transition point. We expect this maximum (minimum) to be the precursor of a peak (dip) close to $\Delta_{c}$ after the quantum dynamics generated following the protocol introduced in Sec.~\ref{sec:Protocol}. To test this, we take as initial state the ground state at large $\Delta_I=2$ and quench $\Delta$ across the neighborhood of $\Delta_{c}$. Due to the high computational cost of the iTEBD calculations for the spin-1 anisotropic Heisenberg chain, we are only able to study dynamics at short times ($t\leq 7$) after the quench. Still, for these short times, Fig.~\ref{Topological}(a) [Fig.~\ref{Topological}(b)] shows that a peak (dip) appears to develop in $dC_1^x/d\Delta$ ($dC_1^z/d\Delta$) about a $\Delta^*$ greater than, but close to, the transition point $\Delta_{c}$. As $t$ increases those peaks sharpen and move toward $\Delta_{c}$. This suggest that our protocol can be used to locate this phase transition.

In the spin-1 anisotropic (XXZ) Heisenberg chain model, like in the ANNNI model, to locate the phase transition using our protocol we rely on the rapid change of local correlations close to the transition point. Next, we study a model in which at the transition point in equilibrium, due to a symmetry, there is a vanishing change in local correlations. The question then is whether this can also be detected in the quantum dynamics and used to locate the transition point.

\begin{figure}[!t]
\includegraphics[width=0.99\linewidth]{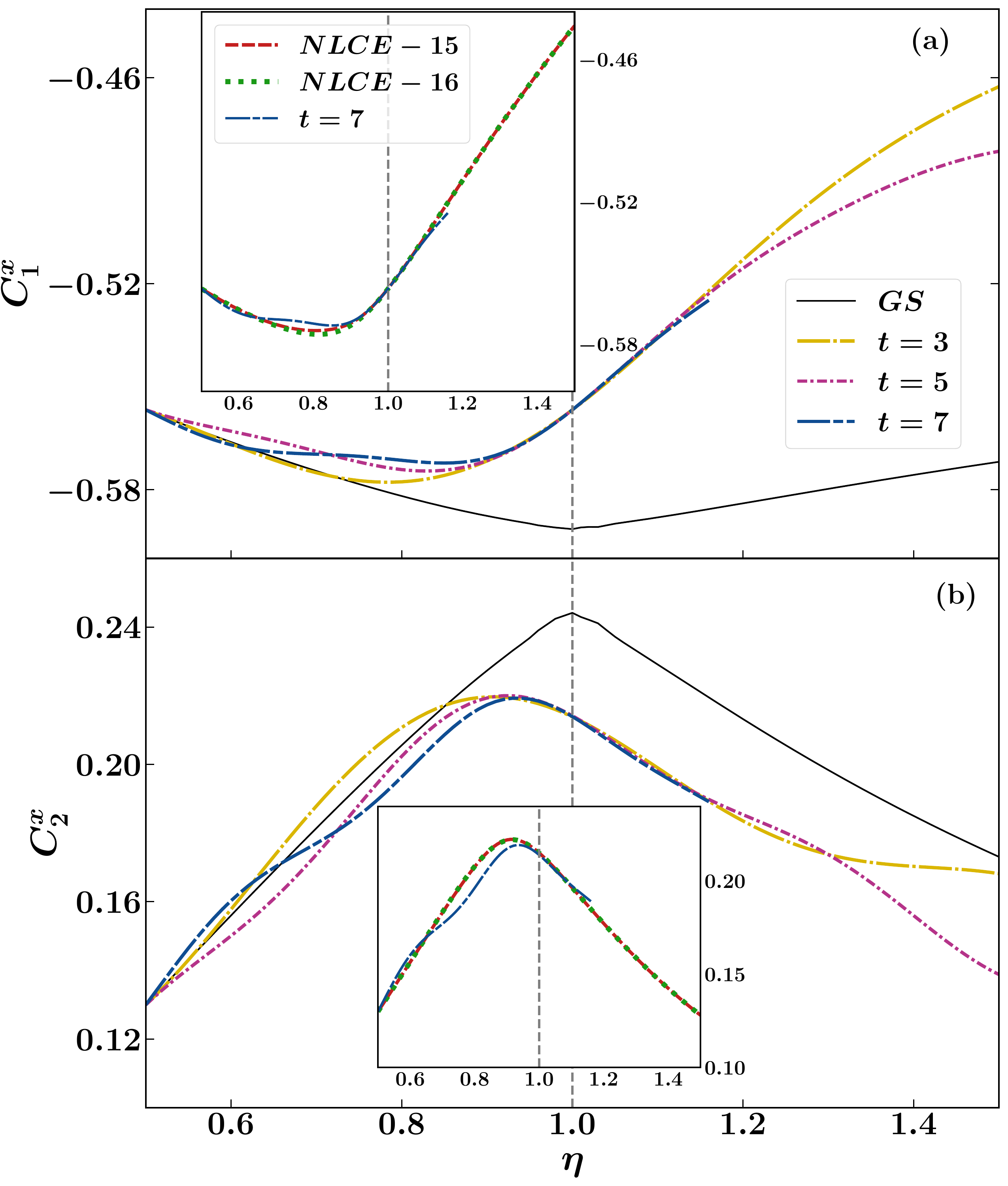}
  \caption{Behavior of local operators, (a) $C_{1}^{x}$ and (b) $C_{2}^{x}$, about the topological transition at $\eta_{c} = 1$ in the bond-alternating Heisenberg chain. The main panels display results of the short-time dynamics (obtained using iTEBD) following quantum quenches, see text for details about the quenches, as well as in the ground state (obtained using iTEBD). The insets show the expected long-time thermal equilibrium results after the quenches evaluated using NLCE to 15 orders (NLCE-15) and 16 orders (NLCE-16), as well as the longest-time result reported in the main panel. All the results exhibit extrema close to the critical point.}
\label{fig:Dimer}
\end{figure}

The model is the bond-alternating Heisenberg model~\cite{Hida_Dimer}
\beq  
\label{H_b-alter}
\hat H \doteq \sum_{i=1}^{L}(\vec{\sigma}_{2i-1}\vec{\sigma}_{2i} + \eta \vec{\sigma}_{2i}\vec{\sigma}_{2i+1}) 
\eeq  
\noindent where $\vec{\sigma_{i}}$ are the Pauli matrices (periodic boundary condition implied). This model exhibits a topological transition between two dimerized phases at $\eta_c = 1$, which can be located using a nonlocal string order parameter. Because of the invariance (up to a rescaling) of Hamiltonian~\eqref{H_b-alter} under $\eta\rightarrow 1/\eta$, one can see that in thermal equilibrium local correlations are symmetric about $\eta_c = 1$. This means that, so long as the correlations depend on $\eta$, they must exhibit a maximum or a minimum at $\eta_c = 1$. In Fig.~\ref{fig:Dimer} we show that this is indeed the case for $C_{1}^{x}$ and $C_{2}^{x}$, defined in Eq.~\eqref{Cx}, in the ground state. At $\eta_c = 1$, $C_{1}^{x}$ exhibits a minimum in Fig.~\ref{fig:Dimer}(a), and $C_{2}^{x}$ exhibits a maximum in Fig.~\ref{fig:Dimer}(b). Next, we explore the fate of those extrema in the quantum dynamics.

We quench the parameter $\eta$ following our protocol in Sec.~\ref{sec:Protocol}, namely, taking the initial state to be the ground state of Hamiltonian~\eqref{H_b-alter} for $\eta^{I}=0.5$ and studying the time evolution under Hamiltonian~\eqref{H_b-alter} with different values of $\eta$. Figure~\ref{fig:Dimer} shows that extrema occur at $\eta^{*}$ close to $\eta_c$, both in the short-time dynamics (studied using iTEBD, and shown in the main panels) and after thermalization (studied using NLCE, and shown in the insets). We note that the position $\eta^{*}$ of the minimum in $C_{1}^{x}$ (maximum in $C_{2}^{x}$) relative to $\eta_{c}$ depends on whether the initial state chosen is the ground state for $\eta^{I}$ greater or smaller than $\eta_{c}$. The minimum (maximum) develops at $\eta^{*} < \eta_{c}$ for $\eta^{I} < \eta_{c}$, as seen in Fig.~\ref{fig:Dimer}(a) [Fig.~\ref{fig:Dimer}(b)], and would develop at $\eta^{*} > \eta_{c}$ for $\eta^{I} > \eta_{c}$. This is also a result of the invariance (up to a rescaling) of Hamiltonian~\eqref{H_b-alter} under $\eta\rightarrow 1/\eta$. Hence, our protocol can also be used in this model for which the transition is located directly using the local observables, as opposed to the earlier models for which it was located using derivatives of the local observables.

\section{Summary and discussion}\label{sec:discussion}

In summary, we have shown that local observables can be used to locate the ferromagnetic to paramagnetic quantum phase transition in the ANNNI chain (a nonintegrable model) both at intermediate times after a quench and at long times after thermalization. The initial states for our quenches were chosen to be ground states of the ANNNI chain deep in the ferromagnetic phase. We explored the effect that changing the magnitude of the quench and starting from initial finite temperature states has in many of our conclusions, and showed that our conclusions are robust against those changes. We also discussed potential experimental tests, as well as the applicability of our protocol to detect topological quantum phase transitions.

More generally, the fact that intermediate-time dynamics, following quenches whose initial states are ground states far from a quantum phase transition, provide a way to locate the quantum phase transition is promising for experiments with ultracold quantum gases~\cite{bloch_dalibard_review_08, esslinger_review_10} and ions~\cite{leibfried_blatt_03, islam_edwards_11}. In those experiments, it is usually straightforward to prepare ground states far away from quantum phase transitions but it is much more challenging to prepare them close to the transitions. The latter is needed to locate the quantum critical point via traditional measurements of the system in equilibrium. Also, not needing to wait long times to observe signatures of the quantum phase transition in the dynamics after the quench is important because, due to heating and other undesirable effects, keeping the dynamics coherent in the experiments becomes increasingly challenging as the evolution time increases. 

\acknowledgements
We thank F. Essler and R. Moessner for useful discussions. 
A. D. thanks S. Bhattacharyya, S. Dasgupta, R. Moessner and S. Roy for earlier collabotaions
in related works. A.D.~and A.H.~acknowledge support from DST-MPI partner group program ``Spin liquids: correlations, dynamics and disorder" between MPI-PKS (Dresden) and IACS (Kolkata), and the visitor's program of MPI-PKS. K.M.~and M.R.~were supported by the National Science Foundation under Grants No.~PHY-1707482 and~PHY-2012145. F.P.~is funded by the European Research Council (ERC) under the European Unions Horizon 2020 research and innovation program (grant agreement No. 771537). F.P.~acknowledges the support of the DFG Research Unit FOR 1807 through grants no.~PO 1370/2-1, TRR80, and the Deutsche Forschungsgemeinschaft (DFG, German Research Foundation) under Germany’s Excellence Strategy EXC-2111- 390814868. This project has also received funding from the European Research Council (ERC) under the European Union’s Horizon 2020 research and innovation programme (grant agreement No. 853443), and M.H.~further acknowledges support by the Deutsche Forschungsgemeinschaft via the Gottfried Wilhelm Leibniz Prize program. This research was also supported in part by the International Centre for Theoretical Sciences (ICTS) during a visit for the program -  Thermalization, Many body localization and Hydrodynamics (Code: ICTS/hydrodynamics2019/11).

\section{Numerical calculations}
\label{Apndx:comp}

\subsection{Infinite Time-Evolving Block Decimation}
\label{Apndx:iTEBD}
In this section, we briefly outline details about the infinite-time evolving block decimation (iTEBD) method. The iTEBD algorithm is based on the infinite matrix-product state (iMPS) representation, which can efficiently represent many-body wave functions with the accuracy controlled by the bond dimension $\chi$ (the error decreases rapidly with increasing $\chi$). A general quantum state $|\Psi\rangle$ on a chain with $L$ sites can be written in the following MPS form~\cite{Vidal_1, Vidal_2, Vidal_3, Schlwck_dmrg_mps}:
\begin{equation}
	|\Psi \rangle = \sum_{s_1, \ldots, s_L}  A[1]^{s_1}A[2]^{s_2} \ldots A[N]^{s_L} | s_1, \ldots ,s_{L} \rangle, \label{eq:mps}
\end{equation}
where, $A[n]^{s_n}$ is a $\chi_{n-1} \times \chi_{n}$ dimensional matrix and $|s_n\rangle$ with $s_n=1,\dots,d$ is a basis of local states at site $n$. For any arbitrary state $|\Psi\rangle$ represented in this product basis $|s_1 \rangle \otimes \ldots \otimes |s_L \rangle$, one can write:
\begin{equation}
	|\Psi \rangle = \sum_{s_1, \ldots, s_L} c_{s_1 \ldots s_L} | s_1, \ldots ,s_{L} \rangle \nonumber.
\end{equation}
Doing repeated Schmidt decomposition on the state $|\Psi \rangle$, one can get the form for the coefficients $c_{s_1 \ldots s_L}$:
\begin{equation}
	c_{s_1 \ldots s_L} = \sum_{s_1, \ldots, s_N}
		\begin{array}{r}
		\Gamma^{[1]s_1} \Lambda^{[1]} \Gamma^{[2]s_2} \Lambda^{[2]} \cdots \Lambda^{[L-1]} \Gamma^{[L]s_L}	
		\end{array} ,
	\label{eq:decomp}
\end{equation}
where $\Gamma's$ are rank-3 tensors, and $\Lambda's$ are positive, real, square-diagonal matrices. After doing the tensor contractions, the structure obtained can be readily identified with a Matrix Product State as in equation~\eqref{eq:mps}.

The size of the tensors $\chi_i$ required to represent a state can be shown to be related to the von Neumann entropy $S_i$ of the
partition $1 \ldots i \colon i+1 \ldots L$, as $S_i \leq 2 \ln \chi_i$. If the entropy is area-law (as is the case for ground states of one-dimensional gapped systems), $\chi_i$ remains finite in the thermodynamic limit.

Using the iTEBD algorithm, one can evaluate the time evolution of a quantum state:
\begin{equation}
	\ket{\psi(t)} = \hat U(t)\ket{\psi(0)},
	\label{eq:tEvol}
\end{equation}
and use the imaginary time evolution $\hat U(\tau) = \exp(-\hat H\tau)$ to find the ground state of the Hamiltonian $\hat H$. Using the Trotter-Suzuki decomposition to the first order, one can write
\begin{equation}
	e^{(\hat A+\hat B)\delta} = e^{\hat A\delta}e^{\hat B\delta} + \mathcal{O}(\delta^2),
	\label{eq:ST1}
\end{equation}
where $\hat A$ and $\hat B$ are operators, and $\delta$ is a small parameter. To use this expression, we write the Hamiltonian as a sum of two-site operators of the form $\hat H=\sum_i \hat h^{[i,i+1]}$ and decompose it as a sum 
\begin{align}
	\hat H &= \hat H_{\rm odd} + \hat H_{\rm even} \notag\\
	&= \sum_{i\; {\rm odd}} \hat h^{[i,i+1]} + \sum_{i\; {\rm even}} \hat h^{[i,i+1]}. 
\end{align}
The terms within one partition act on different sites and thus commute with each other: $[\hat h^{[i,i+1]},\hat h^{[i',i'+1]}] = [\hat h^{[2i-1,2i]},\hat h^{[2i'-1,2i']}] =0$.

One can approximate the time evolution operator for a very small time slice $\delta t\ll 1$, to the first order, using ~\eqref{eq:ST1}, as:
\begin{equation}
	\hat U(\delta t) \approx \left[\prod_{i\; {\rm odd}} \hat U^{[i,i+1]}(\delta t)  \right]\left[\prod_{i\; {\rm even}} \hat U^{[i,i+1]}(\delta t)  \right],
	\label{TimeEvol}
\end{equation}
where
\begin{eqnarray}
	\hat U^{[i,i+1]}(\delta t)=e^{-i \, \delta t \, \hat h^{[i,i+1]}}.
\end{eqnarray}
To determine the suitable $\delta$, one can successively make it smaller to achieve convergence. We used the bond-link dimension $\chi=4000$ to ensure convergence for the longest real-time dynamics results and the time steps used is $\delta = 0.01$. The time evolution in equation~\eqref{eq:tEvol} is obtained by applying the operators $e^{-i \hat H_{\rm odd} \delta}$ and $e^{-i \hat H_{\rm even}\delta}$ iteratively to the initial state $\ket{\psi(0)}$, which has been previously decomposed in the form of an MPS. After the application of each operator at sites $i$ and $i + 1$ the decomposition~\eqref{eq:decomp} is updated, involving at each step only the transformation of the tensors $\Gamma^{[i]}, \lambda^{[i]}$ and $\Gamma^{[i+1]}$ ~\cite{Schlwck_dmrg_mps, Hauschild_Frank_iTEBD}.

For a translational invariant infinite chain, the state can be written in the form of equation~\eqref{eq:decomp}, where $\Gamma^{[i]}$ and $\lambda^{[i]}$ are independent of $i$. Thus, given that the time evolution is generated by two-site operators, only the tensors $\Gamma^{A}$, $\Gamma^{B}$, $\lambda^{A}$, and $\lambda^{B}$ have to be updated, where 
$\Gamma^{A} = \Gamma^{[2i]}, \Gamma^{B} = \Gamma^{[2i+1]}, \lambda^{A}= \lambda^{[2i]}$, and $\lambda^{B} = \lambda^{[2i+1]}$.

In our case, in which we also have a next-nearest neighbor interaction, one can group the sites (merge two neighboring site to one) and proceed with the same algorithm where the local Hamiltonian is now 16$\times$16 instead of 4$\times$4.

\subsection{Numerical linked cluster expansion}
\label{Apndx:NLCE}

For lattice models in the thermodynamic limit $(L\rightarrow\infty)$, NLCE allows one calculate the expectation value of extensive observables $\hat{O}$ per site, $\mathcal{O}=\langle \hat{O}\rangle/L$, as a sum over contributions from all connected clusters $c$ that can be embedded on the lattice:
\begin{equation}\label{nlce_eq}
\mathcal O=\sum_{c}M(c)\times W_{O}(c).
\end{equation}
where $W_{O}(c)$ is the weight of cluster $c$, and $M(c)$ is the number of ways per site in which one can embed $c$ on the lattice. $W_O(c)$ is computed for each cluster $c$ using the inclusion exclusion principle:
\begin{equation}\label{weight_subtraction}
W_{O}(c)=\langle\hat{O}\rangle_c- \sum_{s \subset c} W_{O}(s),
\end{equation}
where $\langle\hat{O}\rangle_c$ is the expectation value of $\hat{O}$ in the cluster $c$, and the sum runs over all connected sub-clusters of $c$. For the smallest cluster $c_0$, $W_{O}(c_0)=\langle\hat{O}\rangle_{c_0}$. 

For each cluster, $\langle\hat{O}\rangle_c = \text{Tr}[\hat{\rho}^c\hat{O}]$, where $\hat{\rho}^c$ is the relevant density matrix in the cluster. For the initial state $\hat{\rho}^c$ is of the form Eq.~\eqref{eq:rhoi}, and for the thermal state used to describe observables after equilibration $\hat{\rho}^c$ is of the form Eq.~\eqref{eq:GE}, with their respective Hamiltonians restricted to the cluster $c$. $\langle\hat{O}\rangle_c$ is calculated numerically using full exact diagonalization. 

We use the maximally connected expansion introduced in Ref.~\cite{rigol_16}, in which each cluster $c$ contains all possible bonds between the sites as per the specific Hamiltonian considered. The order of the NLCE is then the number of lattice sites of the largest cluster $c$ considered in the sum~\eqref{nlce_eq}. The series is convergent when errors in consecutive orders vanish exponentially fast with increasing order.

For the thermal equilibrium results in the bond-alternating Heisenberg model in Sec.~\ref{sec:topological} [Fig.~\ref{fig:Dimer}], we calculate $\langle\hat{O}\rangle_c$ separately for the bond-alternating Hamiltonian and its reflected configuration in $c$ and average them. This extra step is necessary to restore the translational invariance assumed to build the NLCE used. For this model, in order to calculate the temperatures of the thermal equilibrium ensembles used to describe observables after thermalization, we use energies after the quench that are obtained using iTEBD. With those energies, the temperatures are obtained using a 16-order NLCE calculation. The convergence errors in the calculation of the energy are smaller than $5\times 10^{-4}$ for all parameters considered (they are much smaller than $5\times 10^{-4}$ for most parameters considered). 

\section{Transverse-Field Ising Chain}
\label{Apndx:Ising}

The transverse field Ising chain (TFIM) is probably the most studied exactly solvable (integrable) model in the context of quantum phase transitions~\cite{Subir-Book,BKC-Book}. Its Hamiltonian reads
\beq
\hat H \doteq -\sum_{i}^{L}\sigma_{i}^{x}\sigma_{i+1}^{x} -\Gamma\sum_{i}^{L}\sigma_{i}^{z}.
\label{Ham_TFIM}
\eeq
\noindent
It is the noninteracting limit ($\kappa=0$) of our ANNNI Hamiltonian [Eq.~\eqref{H_ANNNI}]. 

\begin{figure}[!b]
\includegraphics[width=0.8\linewidth]{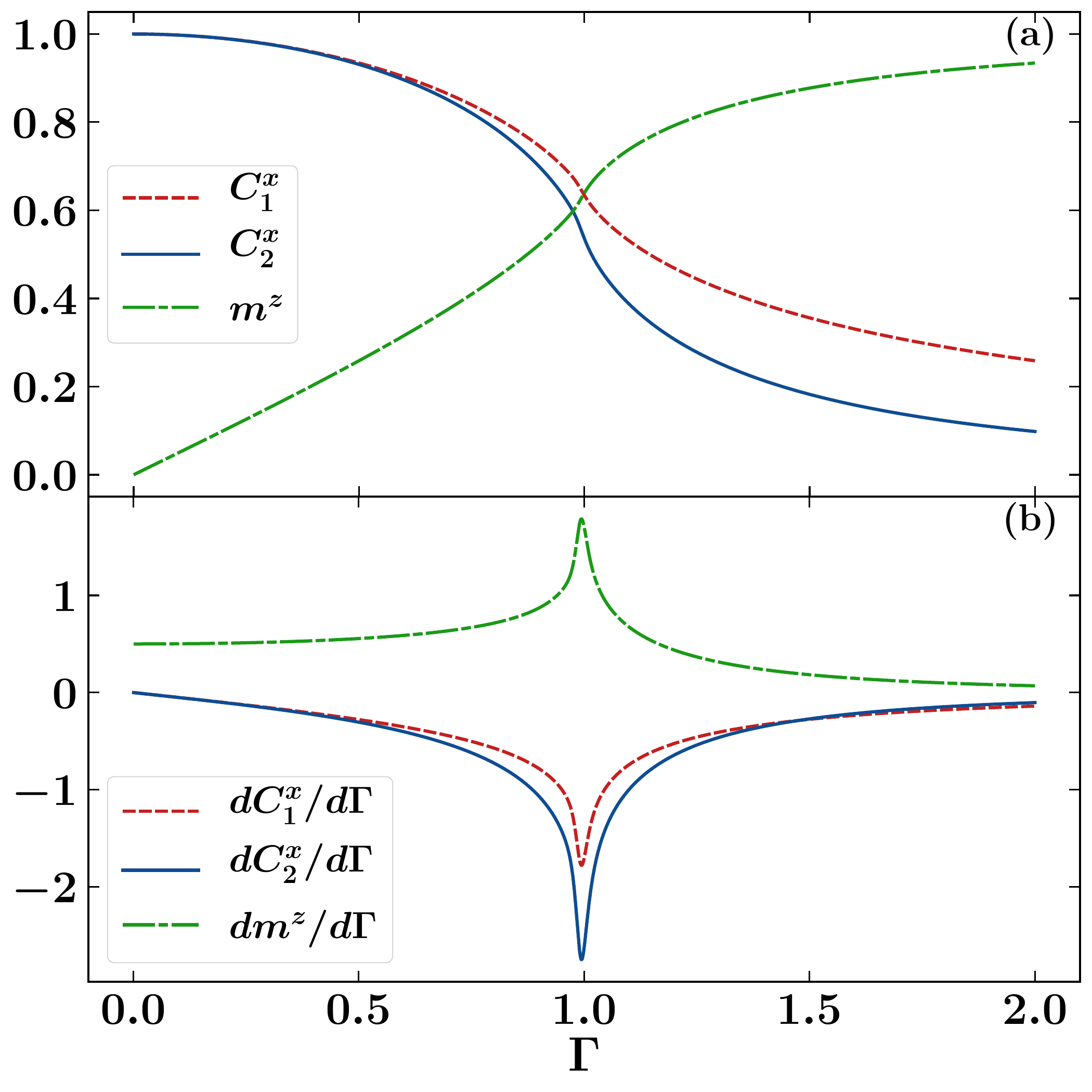}
\caption{Ground-state results for (a) $C_{1(2)}^{x}$, and $m_z$, and (b) their derivatives, as functions of the strength of the transverse magnetic field.}
\label{TFIM_Plots}
\end{figure}

In Fig.~\ref{TFIM_Plots}, we report ground state results for $C_{1(2)}$ and $m_z$ [Fig.~\ref{TFIM_Plots}(a)], and their derivatives [Fig.~\ref{TFIM_Plots}(b)], across the ferromagnetic to paramagnetic phase transition, which occurs in this model at $\Gamma=1$.

\newpage

\bibliography{QPT_HMHPRD}

\end{document}